\RequirePackage[l2tabu,orthodox]{nag}
\documentclass[
    preprintnumbers,
    superscriptaddress,
    final,
    a4paper,
    11pt,
    amsmath,
    amssymb,
    aps,
    nofootinbib
]{revtex4-2}

\usepackage[utf8]{inputenc}
\usepackage[british]{babel}
\usepackage[T1]{fontenc}

\usepackage{microtype}

\usepackage{lmodern}
\usepackage{amsmath}
\usepackage{amsfonts,amsthm,amssymb,mathrsfs}
\usepackage{bm}
\usepackage{graphicx}
\usepackage{float}
\usepackage{booktabs, multirow}
\usepackage[free-standing-units, exponent-product = \cdot, list-units = single, range-units = single]{siunitx}
\usepackage[usenames,svgnames]{xcolor}

\usepackage{csquotes}
\usepackage{hyperref}
\usepackage{cleveref}
\usepackage{xfrac}

\usepackage{acro}
\DeclareAcronym{cc}{
  short=c.c. ,
  long=cosmological constant
}
\DeclareAcronym{nec}{
  short=NEC ,
  long=null energy condition
}
\DeclareAcronym{eom}{
  short=e.o.m. ,
  long=equations of motion
}

\usepackage{tikz}
\usepackage{pgfplots}

\usetikzlibrary{external}
\usetikzlibrary{decorations, arrows, patterns, shapes.symbols, shapes.misc, shadings, decorations.text, backgrounds}
\usepgfplotslibrary{fillbetween}
\usepgfplotslibrary{groupplots}

\pgfplotsset{
  height=0.25\textheight,
  width=0.666\textwidth-2\columnsep,
  grid = both,
  grid style = {opacity = 0.5},
  compat = newest,
  every axis plot/.append style={
    line join=round,
    line cap=round,
    clip=false
  },
}

\usepackage{mathtools}
\usepackage{cancel}
\usepackage{tensor}

\usepackage{tabstackengine}

\usepackage[inline]{showlabels}
\usepackage{soul}
\usepackage{pagecolor}
\usepackage{afterpage}
\usepackage[framemethod=TikZ]{mdframed}

\newcommand{\bigO}{\mathcal{O}}

\newcommand{\Lagrangian}{\mathscr{L}}
\newcommand{\metric}{g}
\newcommand{\ScCurv}{R}
\newcommand{\MPl}{M_{\text{Pl}}}
\newcommand{\MPlSq}{\MPl^2}
\newcommand{\Hubble}{H}
\newcommand{\Hdot}{\dot{\Hubble}}
\newcommand{\Hinput}{\Hubble_{\text{input}}}

\newcommand{\LEH}{\Lagrangian_{\mathrm{E.H.}}}

\newcommand{\Lcc}{\Lagrangian_{\mathrm{c.c. \, relax}}}
\newcommand{\LNECVreh}{\Lagrangian_{\mathrm{NECV+reh}}}

\newcommand{\pii}{\varphi_i}
\newcommand{\pionebg}{\bar{\varphi}_1}
\newcommand{\pione}{\varphi_1}

\newcommand{\pionebgdot}{\dot{\bar{\varphi}}_1}
\newcommand{\pitwo}{\varphi_2}
\newcommand{\pitwodot}{\dot{\varphi}_2}
\newcommand{\pitwodotdot}{\ddot{\varphi}_2}
\newcommand{\pithree}{\varphi_3}
\newcommand{\pithreedot}{\dot{\varphi}_3}
\newcommand{\pithreedotdot}{\ddot{\varphi}_3}

\newcommand{\Xone}{X_1}
\newcommand{\Xtwo}{X_2}
\newcommand{\Xthree}{X_3}

\newcommand{\Vpre}{V_{\text{pre}}}
\newcommand{\Vafter}{V_{\text{after}}}

\newcommand{\KK}{{K}}
\newcommand{\Gthree}{{G}_3}

\newcommand{\PP}{{P}}

\newcommand{\PPX}{\partial_{\Xthree}\PP}
\newcommand{\PPXX}{\partial_{\Xthree\Xthree}\PP}
\newcommand{\PPPiX}{\partial_{\pithree\Xthree}\PP}
\newcommand{\PPPi}{\partial_{\pithree}\PP}
\newcommand{\PPPiPi}{\partial_{\pithree\pithree}\PP}
\newcommand{\PPPiPiX}{\partial_{\pithree\pithree\Xthree}\PP}
\newcommand{\PPPiXX}{\partial_{\pithree\Xthree\Xthree}\PP}
\newcommand{\KKPi}{\partial_{\pitwo}\KK}
\newcommand{\KKPiX}{\partial_{\pitwo\Xtwo}\KK}
\newcommand{\KKX}{\partial_{\Xtwo}\KK}
\newcommand{\KKXX}{\partial_{\Xtwo\Xtwo}\KK}
\newcommand{\KKXChi}{\partial_{\Xtwo\pithree}\KK}
\newcommand{\KKChi}{\partial_{\pithree}\KK}
\newcommand{\KKChiChi}{\partial_{\pithree\pithree}\KK}
\newcommand{\GthreePi}{\partial_{\pitwo}\Gthree}
\newcommand{\GthreePiPi}{\partial_{\pitwo\pitwo}\Gthree}
\newcommand{\GthreePiX}{\partial_{\pitwo\Xtwo}\Gthree}
\newcommand{\GthreePiChi}{\partial_{\pitwo\pithree}\Gthree}
\newcommand{\GthreeX}{\partial_{\Xtwo}\Gthree}
\newcommand{\GthreeXX}{\partial_{\Xtwo\Xtwo}\Gthree}
\newcommand{\GthreeXChi}{\partial_{\Xtwo\pithree}\Gthree}
\newcommand{\GthreeChi}{\partial_{\pithree}\Gthree}
\newcommand{\GthreeChiChi}{\partial_{\pithree\pithree}\Gthree}

\newcommand{\Fone}{F_1}
\newcommand{\Ftwo}{F_2}
\newcommand{\potentialone}{V_1}
\newcommand{\couplingpotential}{V}
\newcommand{\potentialthree}{U}

\newcommand{\alphakick}{\alpha_{\text{kick}}}
\newcommand{\betakick}{\beta_{\text{kick}}}
\newcommand{\betadip}{\beta_{\text{dip}}}
\newcommand{\betakin}{\beta_{\text{kin}}}
\newcommand{\betaI}{\beta_{\text{I}}}

\newcommand{\Hdip}{\Hubble_{\text{dip}}}
\newcommand{\Fkb}{F_{\text{kb}}}
\newcommand{\Fkbzero}{F_{\text{kb,0}}}
\newcommand{\Fkbone}{F_{\text{kb,1}}}
\newcommand{\Tkb}{T_{\text{kb}}}

\newcommand{\Hone}{\Hubble_1}
\newcommand{\Hzero}{\Hubble_0}
\newcommand{\HT}{T}
\newcommand{\Htau}{\tau}
\newcommand{\Tdip}{T_{\text{dip}}}

\newcommand{\soundspeed}{c_s}
\newcommand{\cstwo}{\soundspeed^2}

\newcommand{\ie}{\textit{i.e}}
\newcommand{\eg}{\textit{e.g}}

\newcommand{\Infine}{\textit{In fine}}

\newcommand{\adhoc}{\textit{ad hoc}}

\newcommand{\dummycharacter}{\textcolor{LightGray}{\blacksquare}}

\def\dd{\mathrm{d}}
\def\Mpl{\MPl}
\newcommand{\beal}{\begin{equation}\begin{aligned}}
\newcommand{\enal}{\end{aligned}\end{equation}}

\colorlet{defaultcolor}{Black}
\colorlet{colorpitwo}{Black} %{MediumBlue}
\colorlet{colorpitwodot}{Black} %{DodgerBlue}
\colorlet{colorpithree}{Black} %{ForestGreen}
\colorlet{colorpithreedot}{Black} %{LimeGreen}
\colorlet{colorhubble}{Black} %{DarkViolet}
\colorlet{colorcstwo}{Black} %{DarkOrange}
\colorlet{colorconstraint}{Black} %{HotPink}

\usepackage{verbatim}

\begin{document} %%%%%%%%%%%%%%%%%%%%%%%%%%%%%%%%%%%%%%%%%%%%%%%%%%%%%%%%%%%
%%%%%%%%%%%%%%%%%%%%%%%%%%%%%%%%%%%%%%%%%%%%%%%%%%%%%%%%%%%%%%%%%%%%%%%%%%%%

\preprint{YITP-22-51, IPMU22-0029, RIKEN-iTHEMS-Report-22}

\title{Reheating after relaxation of large cosmological constant}

\author{Paul Martens}
\affiliation{Center for Gravitational Physics and Quantum Information, Yukawa Institute for Theoretical Physics, Kyoto University, 606-8502, Kyoto, Japan}

\author{Shinji Mukohyama}
\affiliation{Center for Gravitational Physics and Quantum Information, Yukawa Institute for Theoretical Physics, Kyoto University, 606-8502, Kyoto, Japan}
\affiliation{Kavli Institute for the Physics and Mathematics of the Universe (WPI), The University of Tokyo, Kashiwa, Chiba 277-8583, Japan}

\author{Ryo Namba}
\affiliation{RIKEN Interdisciplinary Theoretical and Mathematical Sciences (iTHEMS), Wako, Saitama 351-0198, Japan}

\date{\today}

\begin{abstract}
  We present a cosmological model of an early-time scenario that incorporates a relaxation process of the would-be large vacuum energy, followed by a reheating era connecting to the standard hot big bang universe.
  Avoiding fine-tuning the cosmological constant is achieved by the dynamics of a scalar field whose kinetic term is modulated by an inverse power of spacetime curvature \cite{Mukohyama:2003nw,Mukohyama:2003ac}.
  While it is at work against radiative corrections to the dark energy, this mechanism alone would wipe out not only the vacuum energy but also all other matter contents.
  Our present work aims to complete the scenario by exploiting a null-energy-condition violating sector whose energy is eventually transferred to a reheating sector.
  We provide an explicit example of this process and
  thus a concrete scenario of the cosmic onset that realizes the thermal history of the Universe with a negligible cosmological constant.
\end{abstract}

\maketitle

%%%%%%%%%%%%%%%%%%%%%%%%%%%%%%%%%%%%%%%%%%%%%%%%%%%%%%%%%%%%%%%%%%%%%%%%%%%%
%%%%%%%%%%%%%%%%%%%%%%%%%%%%%%%%%%%%%%%%%%%%%%%%%%%%%%%%%%%%%%%%%%%%%%%%%%%%

\section{Introduction}

Observational developments in the past few decades have proven their potential to pin down cosmological parameters to high precision. Despite some recent discrepancy in the determination of the observed value of the expansion rate \cite{Riess:2011,Planck:2018vyg,Bonvin:2016crt,Bernal:2016gxb,Riess:2021jrx}, it is now a well accepted fact that our Universe as a whole not only expands but at an accelerated rate. What causes the accelerated expansion is often dubbed \textit{dark energy}, though there is no agreed consensus about its true nature. The simplest possibility is the \ac{cc}; it can be freely added to the Einstein equation with an arbitrary value at the classical level, while quantum-mechanically the zero-point energy possessed by vacuum fluctuations could contribute to it.

The discrepancy between the observed and theoretically expected values of the \ac{cc}~is the major source of the so-called \ac{cc}~problem.
In the viewpoint of quantum field theory, one would sum all zero-point diagrams and thus obtain some value that is subject to UV physics, which should compose the vacuum energy.
On the other hand, one might wish that the \ac{cc}, written as $\Lambda$, which enters in the Einstein equations, would incorporate all the above contributions.
However, if it does then the observational bounds on the \ac{cc} require an enormous cancellation among different contributions. Without such a cancellation, these two numbers would not match, but more than that, they would differ by many orders of magnitude (the observational bounds being extremely smaller) \cite{Weinberg:1988cp,Martin:2012bt,Padilla:2015aaa}.
The \ac{cc} problem can thus also be seen as a bridging problem between the cosmological and the quantum worlds, and any significant progress on its resolution could eventually lead to a fundamental progress in both paradigms.

Furthermore, the \ac{cc} problem, while already an issue on its own, begs yet another problem: the \textit{coincidence problem}.
It basically asks the following question: why is the small value of the vacuum energy, \ie. the density of dark energy, of the same order as that of dark matter today?
The scaling of the dark energy density with respect to time is different from that of the dark matter, and only a few Hubble times would deviate their values by orders of magnitude.
This implies that we live at a very special moment, nested at the hinge between a dark matter dominated and a dark energy dominated era.
Such a coincidence requires a delicate fine tuning \cite{Velten:2014nra}, 
leaving us with a speculation that there may be a physical mechanism to realize it.

While numerous attempts have been made to solve the issue, including from supersymmetry perspectives to effective modifications of gravity (see \eg. \cite{Weinberg:1988cp,Padilla:2015aaa} for review and references therein), none have yet provided a satisfactory and definitive answer.
One may also be tempted to invoke the so-called \textit{anthropic principle} to give a reasoning to it, \ie.~the dark energy is so small and the transition occurs now, because the physical conditions to our existence are only reunited at this precise moment and with that physical parameter.
While it might as well serve as a valid answer, and one cannot totally refute its relevance,
its nature makes it difficult to test and to quantify new predictions out of it. 
In other words, taking it as the final answer too hastily runs into the risk of missing some more fundamental physical connection, and it therefore appears essential to further investigate a concrete mechanism to address the issue at hand.

Many different paths have been investigated to solve the \ac{cc} problem, and this is not the first time some \ac{cc} relaxation processes, as in our present study, have been considered.
For example, one could consider to have the relaxation unfolds during an unconventional inflationary phase \cite{Rubakov:1999aq}, to work within the framework of some bouncing Universe scenario by dynamically decreasing the value of \ac{cc} during a contracting phase \cite{Graham:2019bfu}, or to forbid a non-vanishing $4$d curvature of a maximally symmetric $3$-brane world-volume enbedded in $5$d spacetime \cite{Arkani-Hamed:2000hpr,Lacombe:2022cbq}). Ideally, one's solution should eventually be linked to some observational signatures (\eg. \cite{Evnin:2020wfw}).
While our approach, which is elaborated below, does not yet expose itself to this last test and rather provides a proof of concept at this stage, it remains within the realm of effective field theories and should still be adaptable for further considerations.

Refs.~\cite{Mukohyama:2003ac,Mukohyama:2003nw} proposed a model that dynamically relaxes the value of the \ac{cc}~to a tiny one thanks to a scalar field
that ever rolls down its potential~\footnote{This scalar field corresponds to $\varphi_1$ in later sections, where the subscript number is introduced to distinguish among three scalar fields that are responsible for different processes in our scenario.}. 
The kinetic term of this field is modulated by a negative power of the Ricci scalar,
and its apparent divergence in the limit of vanishing curvature is dynamically prohibited by the classical motion of the scalar background.
The relaxation mechanism operates as the potential of the scalar field dominates over the kinetic term in the same limit, and yet the dynamics results in a vanishing potential, which is the future attractor of the system.
Importantly, the potential can include not only that of the scalar itself but also all the other (constant) energy contents, that is,
\begin{align}
  \label{eq:1}
  V_{\rm total} = V_{\rm scalar \; alone} + V_{\rm c.c.}  + V_{\rm zero \; point}  + V_{\rm all \; others} \; ,
\end{align}
and what approaches to zero is $V_{\rm total}$; in a way, the scalar field dynamically fixes its potential value $V_{\rm scalar \; alone}$ only to cancel the \ac{cc}, the zero point energy and all the other (constant) contributions. Therefore even if large bare cosmological constant and radiative zero point energy were present, the effective vacuum energy would go as $\Lambda_{\rm eff} \approx V_{\rm total} /\MPlSq \to 0$ after a long time, and the spacetime geometry would anyway approach a flat one.

This relaxation mechanism partially holds a spirit similar to what Weinberg called \enquote{adjustment mechanism} in \cite{Weinberg:1988cp}. The related no-go theorem \cite{Weinberg:1988cp,Oda:2018lgm,Padilla:2015aaa} is evaded in the current mechanism thanks to the fact that, while the theorem only considers the physics at a dynamical equilibrium and thus assumes translational invariance, an essential ingredient of the mechanism considered here is the non-vanishing canonical momentum of the scalar field, rendering the mechanism considered here an exceptional case to the theorem.

However, while this process alone is a powerful mechanism to resolve the \ac{cc}~problem, it also effectively empties the Universe as an inevitable consequence, by eventually diluting everything away:
the total energy in the Universe\,---\,which includes any radiative corrections\,---\,is constrained to asymptotically converge to a null value.
To connect this empty space to our present Universe, the main purpose of this work is to implement a reheating phase after the c.c~relaxation so that the standard hot big bang scenario is revived.
To this aim, we not only construct a stable model that achieves the scenario, but also demonstrate it by exhibiting a concrete realisation along a numerical verification.

For the reheating phase, we rely in part on the Lagrangian of the Horndeski theory \cite{Kobayashi:2011nu,Horndeski:1974wa,Deffayet:2011gz}, as using the latter has been shown to allow to violate the \ac{nec} in a stable manner \cite{Rubakov:2014jja,Nishi:2016wty}.
The gravitating energy elevated by the \ac{nec} violation is then transferred to another sector that eventually reheats the Universe. In order not to disrupt the \ac{cc}~relaxation mechanism, which is always in action, the \ac{nec}-violating and reheating phases must occur for sufficiently short time compared to the time scale of the relaxation. Moreover, we assume that these phases take place periodically, and our observed Universe can be the result from any one of those recurring occurrences.
\Infine, the dynamical cosmological constant relaxation described by our model would thus allows us to avoid invoking the anthropic principle to solve the \ac{cc} problem.

To present our model, this work shall first describe the overall cosmic history our model suggests and take this occasion to review the cosmological relaxation process we employ (\cref{sec:Overall_picture_of_the_cosmic_history}).
The \ac{nec}-violating and the reheating sectors are then assessed from a theoretical perspective by considering their linear perturbations, explicitly providing the conditions under which we avoid having any ghost or gradient instability against the background dynamics of the desired behavior (\cref{sec:Analysis_of_perturbative_stability}).
We shall then focus on the reheating process, and the model and all its composing functions are now either explicitly chosen or deduced (\cref{sec:A_concrete_implementation}).
Lastly, we use a numerical approach to explicitly and qualitatively witness the whole reheating process unfold (\cref{sec:Numerical_approach}), before concluding our study (\cref{sec:Conclusion_and_outlook}).

%%%%%%%%%%%%%%%%%%%%%%%%%%%%%%%%%%%%%%%%%%%%%%%%%%%%%%%%%%%%%%%%%%%%%%%%%%%%
%%%%%%%%%%%%%%%%%%%%%%%%%%%%%%%%%%%%%%%%%%%%%%%%%%%%%%%%%%%%%%%%%%%%%%%%%%%%

\section{Overall picture of the cosmic history}
\label{sec:Overall_picture_of_the_cosmic_history}

%%%%%%%%%%%%%%%%%%%%%%%%
%%%%%%%%%%%%%%%%%%%%%%%%
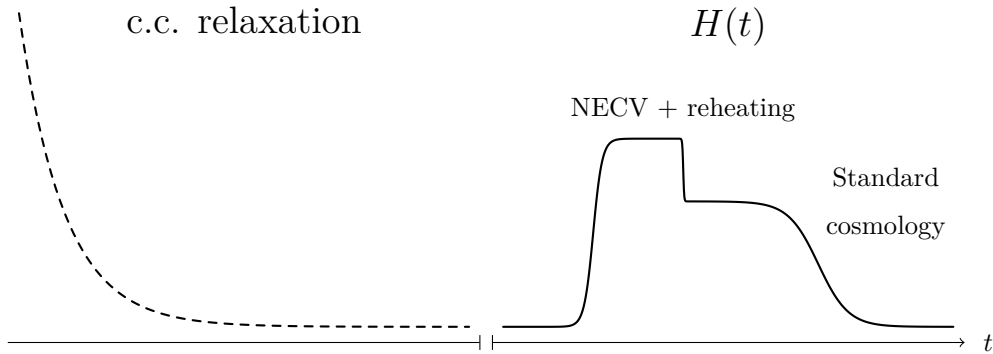
\begin{figure}
  \centering
  \tikzsetnextfilename{schematic_cosmic_evolution}
\begin{tikzpicture}[]%outer sep=1ex
    \begin{groupplot}[
        group style={
            group size = 2 by 1,
            horizontal sep = 1ex,
        },
        width={0.49\textwidth},
        height={0.25\textheight},
        grid=none,
    ]
        \nextgroupplot[
            x tick style={red,opacity=0,},xtick align = center,xticklabels={,,},ymajorticks=false, % ticks=none, % (i) All this line is to have some small padding at the bottom
            axis x line=bottom,
            y axis line style={draw=none},
            ymin=-0.05,
            ymax=1.05,
            xmin=-0.25,
            xmax=10.25,
            axis line style={-|, thin,},
        ]
            \addplot[
                domain=0:10,
                samples=100,
                color=Black,
                dashed,
                thick,
            ] {
                exp(-x)
                % + 0.6/(1+exp(-10*(x-12)))
                % - 0.2/(1+exp(-100*(x-14)))
                % - 0.4/(1+exp(-3*(x-17)))
              };
            \node[below] at (rel axis cs:0.5, 1) {\Large c.c. relaxation};
        \nextgroupplot[
            ticks=none,
            axis x line=bottom,
            y axis line style={draw=none},
            ymin=-0.05,
            ymax=1.05,
            xmin=9.75,
            xmax=20.25,
            axis line style={|->, thin,},
            xlabel=$t$,
            every axis x label/.style={
                % style={font=\Large},
                at={(ticklabel* cs:1.02)},
                anchor=west,
            },
        ]
            \addplot[
                solid,
                thick,
                color=Black,
                domain=10:20,
                samples=1000,
            ] {
                % exp(-x)
                + 0.6/(1+exp(-10*(x-12)))
                - 0.2/(1+exp(-100*(x-14)))
                - 0.4/(1+exp(-3*(x-17)))
              };
            % \draw[red] (rel axis cs:0,0.5) -- (rel axis cs:0,-0.1);
            \node[above] at (axis cs:14,0.625) {NECV + reheating};
            \node[above, align=center] at (axis cs:18.5, 0.25) {Standard\\ cosmology};
            \node[below] at (rel axis cs:0.5, 1) {\Large $\Hubble(t)$};
    \end{groupplot}
\end{tikzpicture}
  \caption{Schematic illustration of the cosmic evolution according to the scenario studied in this paper. The dashed curve on the left corresponds to the initial phase of the \ac{cc} relaxation, and the solid curve on the right depicts the \ac{nec} violation and the reheating dynamics. The Universe is eventually filled with radiation and recovers the standard cosmic history. This figure is highly schematic and not to scale.}
  \label{fig:Hevol_entire}
\end{figure}
%%%%%%%%%%%%%%%%%%%%%%%%
%%%%%%%%%%%%%%%%%%%%%%%%

The vacuum energy that drives the accelerated cosmic expansion at the present time is, whatever its true nature is, observed to have an extremely small value, and this smallness demands an explanation. We explore a dynamical solution to this issue, and in this regard, we employ the mechanism originally proposed in \cite{Mukohyama:2003ac,Mukohyama:2003nw}.
However this mechanism alone leads to an empty universe, that is, it not only reduces the contribution from the \ac{cc}~but also dilutes all other contents of the Universe.
In order to incorporate an additional process to eventually populate the Universe with energetic radiation,
a couple of potential candidates 
have been sketched in \cite{Alberte:2016izw}. Our mechanism in this paper shares some aspects with their \enquote{fast violation} section but extends it. We later provide a concrete realization of the reheating and show with numerical analyses that the scenario can indeed be achieved without any instability.

To this end, the mechanism we propose to incorporate three essential phases through the cosmic history, namely, in order:
\begin{enumerate}
    \item \textbf{Cosmological constant relaxation.} A large value of the \ac{cc}, together with all the other matter content, is dynamically relaxed to the small present value [\cref{subsec:relax} for summary and \cite{Mukohyama:2003nw,Mukohyama:2003ac} for details].
    \item \textbf{Null energy condition violation.} The Universe is energetically revived by accommodating a phase violating the \ac{nec} [\cref{subsec:NECreheat} for summary and \cref{sec:A_concrete_implementation} for details].
    \item \textbf{Reheating.} The Universe reheats and connects this process to the standard cosmological picture [\cref{subsec:NECreheat} for summary and \cref{sec:A_concrete_implementation} for details].
\end{enumerate}
This overall picture is schematically illustrated in \cref{fig:Hevol_entire}.
After the relaxation, the \ac{nec} violation re-energize the Universe, which is the first rise of $\Hubble$ in the figure; after some time, $\Hubble$ quickly drops to some non-zero value, corresponding to the end of the \ac{nec}-violating phase during which a fraction of energy is transferred to the reheating sector.
Eventually this sector decays to radiation, recovering the standard big bang cosmic history.
We assume that the potential for the \ac{nec}-violating sector is periodic so that the sequence described above (the \ac{cc} relaxation, \ac{nec} violation and reheating) repeats many times until the \ac{cc} actually goes down to the observed tiny value.

The model setup can thus be described by the action
\begin{equation}
    \label{action_total}
    S = \int \dd^4x \sqrt{-g} \, \Big(
        \LEH[g]
        + \Lcc[\pione, g]
        + \LNECVreh[\pitwo, \pithree, g]
    \Big) \; ,
\end{equation}
where $\LEH$, $\Lcc$ and $\LNECVreh$ describe the Einstein-Hilbert part, the \ac{cc} relaxation mechanism, and the combined sector of \ac{nec} violation and reheating, respectively. The fields $\pione$, $\pitwo$ and $\pithree$ are responsible for the \ac{cc}~relaxation, \ac{nec} violation and reheating, respectively, and $g$ denotes the spacetime metric.
Let us note that, as we see below and  concretely show in \cref{sec:A_concrete_implementation}, the reheating sector is nontrivially coupled to the \ac{nec}-violating sector, and thus they are not separable at the level of Lagrangian.
In the following subsections, we summarize the gist of the \ac{cc}~relaxation, \ac{nec} violation and reheating sectors individually and the requirements to achieve the desired history of the Universe.

\subsection{Cosmological constant relaxation}
\label{subsec:relax}

We follow \cite{Mukohyama:2003ac,Mukohyama:2003nw} to drive an initially large \ac{cc} to the tiny value observed today, and this subsection serves as a brief review of the mechanism.
This relaxation mechanism is driven by a scalar field $\pione$ that is non-minimally coupled to gravity, and the corresponding Lagrangian $\Lcc$ together with the Einstein-Hilbert part $\mathscr{L}_{\rm E.H.}$ introduced in \cref{action_total} reads
\begin{equation}
    \label{eq:cosmological_constant_lagrangian}
    \Lagrangian = 
        \rlap{$\overbrace{
            \underbrace{
                \frac{\MPlSq}{\phantom{f}2\phantom{f}} \ScCurv
            }_{\LEH}
            \phantom{ + \alpha \ScCurv^2 }
        }^{ \text{gravity}}$}
        \frac{\MPlSq}{2} \ScCurv
        +
        \underbrace{
            \alpha \ScCurv^2
            + \frac{\Xone}{f(\ScCurv)}
            - \potentialone(\pione)
        }_{\Lcc} \; ,
\end{equation}
where we hereafter denote the kinetic terms of scalar fields $\varphi_i$ ($i=1,2,3$) by
\begin{equation}
  X_i \equiv - \frac{1}{2} \, g^{\mu\nu} \partial_\mu \pii \partial_\nu \pii \; ,
\end{equation}
and where $R$ is the Ricci scalar of the spacetime metric $\metric_{\mu\nu}$, $\potentialone$ the potential of $\pione$, and $\MPl$ and $\alpha$ are the (normalized) reduced Planck mass and a dimensionless constant, respectively.
The curvature quadratic term $\alpha \ScCurv^2$ with \eg. $\alpha=\mathcal{O}(1)$ ($>0$) is needed to tame the instability that would otherwise arise during the course of relaxation. The term is purely gravitational and thus could in principle be combined with the Einstein-Hilbert term $\LEH$ to together compose a gravitational action.
Nevertheless, since it only affects the stability during the relaxation and becomes negligible at later stages of our scenario, we simply include it in $\Lcc$.%
\footnote{Other higher-order terms of curvature invariants such as $R_{\mu\nu} R^{\mu\nu}$ and $R_{\mu\nu\rho\sigma} R^{\mu\nu\rho\sigma}$ should arise due to quantum corrections. 
We can always rearrange a linear combination of $R^2$, $R_{\mu\nu} R^{\mu\nu}$ and $R_{\mu\nu\rho\sigma} R^{\mu\nu\rho\sigma}$ into another one of $R^2$, the Gauss-Bonnet term and the Weyl tensor squared. Since the Gauss-Bonnet term does not contribute to equations of motion, and the Weyl squared term is responsible for the ghost modes in UV, we call the coefficient of $R^2$ in this linear combination as $\alpha$ appearing in eq.~\eqref{eq:cosmological_constant_lagrangian}.
Unless fine-tuned, the coefficients of those higher curvature terms are expected to be of $\mathcal{O}(1)$ in the units of $\MPl$, and the would-be ghost modes associated with them have masses of order $\MPl$. 
Therefore the higher-order terms are irrelevant at energies and momenta sufficiently below the Planck scale, with which we are concerned, and the stability only requires $\alpha >0$ in IR.}
Note that any non-zero \ac{cc}~term, including the vacuum energy originated from the quantum fluctuations of matter fields, can be absorbed into $\potentialone(\pione)$ without loss of generality.

A crucial part for the mechanism to work is that the coefficient of the kinetic term of $\pione$ diverges in the limit of vanishing $R$.
To this end, we demand that $f$ vanishes at $\ScCurv = 0$ as
\begin{align}
  f(\ScCurv) \approx \left( \frac{\ScCurv^2}{\MPl^4} \right)^{m} \; ,
\end{align}
where $m$ is some positive number and we assume $m>3/2$ for the reason that we explain later in this subsection. The constant $\MPl^4$ is introduced to make the function dimensionless, and the overall normalization of $f$ can always be absorbed by redefining $\Xone$. Since the kinetic term depends nonlinearly on the curvature, the system described by the Lagrangian \eqref{eq:cosmological_constant_lagrangian} contains two scalar degrees of freedom. It is straightforward to show that both scalar degrees satisfy the no-ghost condition and that the two speeds of propagation are both unity in the low-energy regime~\footnote{See \eg. section~V.A.4 of \cite{DeFelice:2010gb}.}.

Starting with the kinetic Lagrangian of the assumed form $\Xone/f(R)$, quantum corrections may generate a more general kinetic Lagrangian consisting of terms of the form $\Xone^{q_i}/(\ScCurv^2/\MPl^4)^{m_i}$ ($i=1,2,\cdots$). In this case, what controls the \ac{cc} relaxation is the most singular-looking term. Fortunately, the more singular-looking the dominant term is, the more robust the \ac{cc} relaxation mechanism is. Therefore, less singular-looking terms generated by quantum corrections do not spoil the \ac{cc} relaxation mechanism, while more singular-looking terms generated by quantum correction simply strengthen it. See \cite{Mukohyama:2003ac} for some details. In the rest of the present paper, for simplicity we consider the simplest kinetic Lagrangian consisting of a term with $q_i=1$ and $m_i=m>3/2$. 

Another important ingredient is that the potential $\potentialone(\pione)$ crosses zero at some finite value of $\pione$. Starting the evolution from a positive value of $\potentialone$, $\pione$ rolls down the potential and approaches $0$, around which $\potentialone$ can be well approximated by a linear form
\begin{align}
  \label{V_form}
  \potentialone(\pione) \simeq c \MPl^3
  \left(
  \pione - v
  \right) \; ,
\end{align}
where $v$ is the value of $\pione$ at which the potential would cross $0$ and $c$ is some dimensionless constant.
Then, on the flat Friedmann-Lema\^itre-Robertson-Walker (FLRW) background, the equation of motion for the homogeneous background of $\pione$, denoted by $\pionebg$, takes the form
\begin{align}
  \label{EOM_phi1}
  \frac{\partial \Pi_1}{\partial \mathcal{N}} +
  \left(
  3 + \epsilon
  \right) \Pi_1
  + c = 0 \; ,
\end{align}
where $\Hubble$ is the Hubble expansion rate, $\mathcal{N} \equiv \ln a$ is the number of e-folds with $a$ being the scale factor, $\Pi_1 \equiv \Hubble^2 \partial_{\mathcal{N}} \pionebg / ( \Mpl^3 f)$, and $\epsilon \equiv - \partial_{\mathcal{N}} \Hubble / \Hubble$.
Eq.~\eqref{EOM_phi1} makes it evident that, for a negligible time variation of $\epsilon$, the stationary solution is given by $\Pi_1 \simeq - c \left( 3 + \epsilon \right)^{-1}$.

At late time the Friedmann equation around the stationary solution takes the approximate form, that is,
\begin{equation}
  \label{eq:V_from_eq_7_from_Shinjis_paper}
  \potentialone \simeq 3 \Mpl^2 \Hubble^2 \; .
\end{equation}
This equation, together with the aforementioned result, leads to the time variation of $\potentialone$ as
\begin{align}
  \label{Vvariation}
  \frac{\partial}{\partial \mathcal{N}}
  \left(
  \frac{2 \potentialone}{\Mpl^4}
  \right)^{2-2m}
  \simeq 24 \, c^2
  \left(
  m - 1
  \right)
  \frac{
  \left(
  2 - \epsilon
  \right)^{2m}}{3+\epsilon}
  \simeq 2^{2m+3} \, c^2 \left( m - 1 \right)
  \; ,
\end{align}
where in the last equality the fact that $\epsilon$ is small during the relaxation phase is used.
This clearly shows that, when $\potentialone$ approaches to $0$, the field $\pionebg$ in fact stalls, and $\potentialone$ never crosses $0$, namely
\begin{align}
  \label{eq:V1_approaches_0_when_N_goes_to_infinity}
  \potentialone \to + 0 \; \qquad
  \mbox{ as } \quad
  \mathcal{N} \to +\infty
\end{align}
is the asymptotic behavior, provided that $m>3/2$ as we have assumed.
In fact, the behavior \eqref{eq:V1_approaches_0_when_N_goes_to_infinity} only requires $m>1$ besides the smallness of $\epsilon$. However, if the scalar kinetic term $\Xone /f$ in \cref{eq:cosmological_constant_lagrangian} dominated over the $\alpha \ScCurv^2$ term at low energy ($\Hubble \ll \Mpl$), the dynamics of the system would destabilize the stationary solution above \cite{Mukohyama:2003ac}. To prevent this, we demand $\Xone /f < \alpha \ScCurv^2$ at low energy, which can be achieved for $m > 3/2$ and is self-consistent with the solution obtained above.

The essence of the mechanism to relax a large \ac{cc} is as described above. It is worth stressing that the most important ingredient is the coefficient of $\Xone$ in \cref{eq:cosmological_constant_lagrangian} that has a singular-looking form in the limit $\ScCurv \to 0$, and this mechanism is effective as long as the most singular term among many other possible ones has the behavior described here.
While quantum corrections should produce additional regular operators in the action, the motion of $\pionebg$ nonetheless drives the total potential to the vanishing value.
On the other hand, one can show that quantum corrections do not generate a potential that is singular at $\ScCurv=0$. Singular-looking kinetic terms of the form $\Xone^{q_i}/(\ScCurv/\MPlSq)^{2m_i}$ ($i=1,2,\cdots$) can be generated by quantum correction but this does not cause a problem. Actually, the more singular-looking the radiatively-corrected kinetic term is, the more robust the \ac{cc} relaxation mechanism becomes. 

Now, we have achieved a tiny value of the \ac{cc}/vacuum energy after a sufficiently long time. However, by that same achievement, since, under the condition $m>3/2$, the potential $\potentialone$ approaches zero more slowly than matter and radiation, the Universe would be empty after the mechanism under consideration takes place.
The Universe thus needs to be \enquote{reheated} once the \ac{cc} is driven to a small value.
This is the subject of the subsequent subsections, and is the main purpose of our present study.

\subsection{Null energy condition violation and reheating}
\label{subsec:NECreheat}

After the mechanism in \cref{subsec:relax} operates, not only the \ac{cc}~and vacuum energy but also all other energy contents decrease to a negligible value. In order to connect to the known cosmic thermal history, \enquote{reheating} thus needs to subsequently take place to re-populate the Universe with energetic radiation.
The \ac{nec} is necessarily violated to achieve this scenario, in order for the energy required for reheating to be temporarily available.
Another field that mediates the reheating process is destabilised through its coupling to the \ac{nec}-violating sector, and its  acquired energy is finally transferred to radiation.

In order to stably violate the \ac{nec} we employ a subclass of the Horndeski theory \cite{Horndeski:1974wa,Deffayet:2011gz,Kobayashi:2011nu}, whose scalar field is now denoted by $\pitwo$. The reheating field $\pithree$ couples to this sector, and we introduce direct couplings besides the gravitational one for efficient energy transfer.
A minimal setup that satisfies these requirements adopts the following form of the Lagrangian in \cref{action_total},
\begin{align}
  \label{eq:NECVreheat}
  \LNECVreh = \KK(\pitwo , \Xtwo, \pithree) - \Gthree(\pitwo , \Xtwo, \pithree) \, \Box \pitwo  + \PP(\pithree , \Xthree) \; ,
\end{align}
where $\pitwo$ invokes the \ac{nec} violation.
Here $\KK$ and $\Gthree$ are some functions of $\pitwo$, $\Xtwo$ and $\pithree$, while $\PP$ is a function of $\pithree$ and $\Xthree$ only.
The reheating field $\pithree$ is implicitly coupled to radiation.
For simplicity we assume that $\LNECVreh$ is in the Einstein frame and hence the total Lagrangian that is relevant to the present mechanism is $\LEH + \LNECVreh$.
As far as $\pione$ stays almost constant, this assumption is expected to be valid since, as explicitly shown in \cite{Mukohyama:2003ac} for linear perturbations, the model \eqref{eq:cosmological_constant_lagrangian} recovers general relativity at low energy.
For the consistency of this treatment, we shall later clarify what we precisely mean by being \enquote{almost constant} and obtain the condition under which $\pione$ stays almost constant during the \ac{nec} violation and reheating.

%%%%%%%%%%%%%%%%%%%%%%%
%%%%%%%%%%%%%%%%%%%%%%%
\begin{figure}
  \centering
  \tikzsetnextfilename{schematic_Hubble_history}
\begin{tikzpicture}
    \begin{axis}[
        width={0.6\textwidth},
        height={0.25\textheight},
        grid=none,
        x tick style={red,opacity=0,},xtick align = center,xticklabels={,,},ymajorticks=false, % ticks=none, % (i) All this line is to have some small padding at the bottom
        axis x line=bottom,
        y axis line style={draw=none},
        ymin=-0.05,
        ymax=1.05,
        xmin=9.75,
        xmax=20.25,
        axis line style={->, thin,},
        xlabel=$t$,
        every axis x label/.style={
            % style={font=\Large},
            at={(ticklabel* cs:1.02)},
            anchor=west,
        },
    ]
            \addplot[
                solid,
                thick,
                color=Black,
                domain=10:20,
                samples=1000,
            ] {%exp(-x)
                + 1.0/(1+exp(-4.5*(x-13)))
                - 1.0/(1+exp(-100*(x-17)))
              };
            \node[left] at (rel axis cs:1.0, 0.8) {\Large $\Hubble(t)$};
    \end{axis}
\end{tikzpicture}
  \caption{Schematic illustration for the Hubble history, with $\Hubble$ the Hubble expansion rate. This figure is simplified in that it only extracts the behavior of the target \ac{nec} violation, with all the other sectors, including the \ac{cc}~relaxation and reheating sector, turned off.}
  \label{fig:Hevol_NECV}
\end{figure}
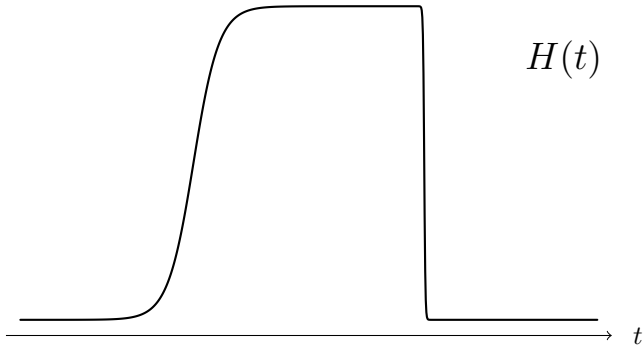
%%%%%%%%%%%%%%%%%%%%%%%
%%%%%%%%%%%%%%%%%%%%%%%
%
The \ac{cc}~relaxation sector discussed in \cref{subsec:relax} is decoupled from the rest of the physics except through gravity. The relaxation mechanism continues to operate throughout the cosmic history to keep the vacuum energy from overdominating the Universe.
Since it takes many (current) Hubble times to make the \ac{cc}~small enough, we assume that the functions $\KK$ and $\Gthree$ are periodic in the \ac{nec}-violating field $\pitwo$ so that a \ac{nec}-violating phase occurs periodically and (partial) reheating may occur several times, in order to avoid a miraculous fine-tuning with respect to the timing of reheating. On top of the periodicity, we assume that the functions of $\pitwo$ enjoy approximate shift symmetry for most of its domain and that the \ac{nec} violation (and reheating) is restricted to a short duration where the approximate shift symmetry is broken.

For the purpose of presentation, let us for the moment turn off the reheating sector $\pithree$ in order to focus on the \ac{nec} violation part. The Lagrangian of \cref{eq:NECVreheat} is then reduced to
\begin{align}
  \mathcal{L}_{\mathrm{NECV}} \equiv \mathcal{L}_{\mathrm{NECV + reheat}} \big\vert_{\pithree = 0, \; P=0} = \tilde \KK (\pitwo, \Xtwo) - \tilde \Gthree ( \pitwo , \Xtwo ) \,
  \Box \pitwo  \; ,
\end{align}
where $\tilde \KK \equiv \KK \vert_{\pithree = 0}$ and $\tilde \Gthree \equiv \Gthree \vert_{\pithree = 0}$. For simplicity we adopt the following ansatz for $\tilde \KK$ and $\Gthree$, keeping only terms lower-order in $\Xtwo$,  
\begin{align}
  \label{eq:KK_and_G3_with_pithree_turned_off}
  \tilde \KK = \tilde f_1 (\pitwo) \, \Xtwo + \tilde f_2 (\pitwo) \, \Xtwo^2 - \tilde V(\pitwo) \; , \qquad
  \tilde \Gthree = \tilde f (\pitwo) \, \Xtwo \; .
\end{align}
The functions $\tilde{f}_1$, $\tilde{f}_2$ and $\tilde{f}$ are periodic in $\pione$ with a common period and should have appropriate forms -- their respective forms we consider in the present study, especially for the purpose of numerical evaluations, are fixed according to the desired evolution of the Hubble expansion rate $\Hubble$ and are reconstructed using the equations of motion. This procedure is discussed in detail in \cref{subsec:reconst_necv}.
With these potential and kinetic structures, the motion of $\pitwo$ accomplishes to violate the \ac{nec} and increases the value of $\Hubble$ for some finite time.
A schematic shape of the resultant evolution of $\Hubble$ with the \ac{nec} violation is illustrated as a function of time in \cref{fig:Hevol_NECV}.
In this figure, all the matter contents other than the \ac{nec}-violating $\pitwo$ are excluded. Once the reheating sector $\pithree$ couples to $\pitwo$, the behavior of $\Hubble$ is modified to the one captured in \cref{fig:Hevol_entire}.

In order to reheat the Universe, the reheating field $\pithree$ transfers the energy acquired during the \ac{nec}-violating period to radiation. It thus couples to $\pitwo$, and this is realized by reintroducing the $\pithree$ dependence into the previous functions. Formally, we make the changes
\begin{align}
  \label{eq:Reintroducing_pithree_in_F1F2}
  \tilde{f}_{1,2}(\pitwo) \; \to \; f_{1,2} (\pitwo, \pithree) \; , \qquad
  \tilde{f} (\pitwo) \; \to \; f(\pitwo, \pithree) \; , \qquad
  \tilde{V} (\pitwo) \; \to \; V(\pitwo, \varphi_3) \; ,
\end{align}
and reintroduce $P(\pithree, \Xthree)$. For simplicity we assume that $\pithree$ is a canonically normalized scalar with a simple potential, except for its coupling to $\pitwo$ via \eqref{eq:Reintroducing_pithree_in_F1F2}. In other words, $P$, in \cref{eq:NECVreheat}, takes the form
\begin{align}
  \label{eq:General_form_of_P_for_pithree}
  P(\pithree , \Xthree) = \Xthree - U(\pithree) \; .
\end{align}
Thanks to these couplings in \eqref{eq:Reintroducing_pithree_in_F1F2}, the effective potential of $\pithree$ basically changes over time with respect to the motion of $\pitwo$. As already invoked, we assume that $\pitwo$ enjoys approximate shift symmetry for most of its domain except for the short \ac{nec} violating period. This in particular means that the effective potential for $\pithree$ changes only in the vicinity of the \ac{nec} violating period. We choose the coupling between $\pitwo$ and $\pithree$ so that, once the \ac{nec} violation starts operating, the effective potential forms a new local minimum in which $\pithree$ gets trapped. After the \ac{nec}-violating phase ends, $\pithree$ stays at a new minimum for a while, but, after a finite time, it eventually rolls back to the initial true minimum of the potential, around which it oscillates and thus drives the reheating.
For illustration of these behaviors, we refer to \cref{fig:Reheating_potential_schematic}.
These phase transitions in $\pithree$ triggered by $\pitwo$ are the essential ingredient of our reheating mechanism.

Due to the approximate shift symmetry in the $\pitwo$ direction, the system away from the \ac{nec} violation (and reheating) period realizes a phase of ghost condensation~\cite{Arkani-Hamed:2003pdi,Arkani-Hamed:2003juy} and the stress-energy tensor from $\mathcal{L}_{\mathrm{NECV + reheat}}$ acts as an additional contribution to the \ac{cc}. Without loss of generality, this additional contribution to the \ac{cc} can be absorbed into $\potentialone(\pione)$ so that the \ac{cc} away from the \ac{nec} violation (and reheating) period is precisely $\potentialone/\MPlSq$.

While this process of the \ac{nec} violation and reheating is overall decoupled from the first scalar field $\pione$, one still has to be careful of the risk of overshooting.
Indeed, the \ac{nec} violating and reheating eras break the simple relation \eqref{eq:V_from_eq_7_from_Shinjis_paper} between the potential $\potentialone$ for $\varphi_1$ and the spacetime curvature, the latter of which controls the relation between $\Pi_1$ and $\dot{\bar\varphi}_1$. During the \ac{nec} violation and reheating eras, the equation \eqref{EOM_phi1} for $\Pi_1$ is unchanged and thus $\Pi_1 = - c \times \mathcal{O}(1)$ still holds. On the other hand, $f(R)$ suddenly increases and affects the value of $\pionebgdot = \Mpl^3\Hubble^{-1} f \Pi_1$. As a result, the field $\pione$ rolls down its potential $\potentialone$ faster and may overshoot the zero of $\potentialone$. Because of this risk, the \ac{nec}-violating part should not last indefinitely; it needs to be constrained in time.

To better understand the issue at hand, let us consider the following calculations.
Close to the zero of its potential $\potentialone$, the field $\pione$ follows a smooth evolution.
Before the \ac{nec} violation and reheating, $\potentialone$ is (virtually) locked at a height of say $\Vpre > 0$.
To avoid a runaway down the potential into effectively negative \ac{cc}, the potential after the \ac{nec} violation and reheating, denoted as $\Vafter \equiv \Vpre+\Delta\potentialone$, must stay above zero. We thus require
\begin{equation}
  \label{cond_overshoot}
  |\Delta \potentialone| < \Vpre \; 
\end{equation}
during the \ac{nec} violation and reheating. One can easily estimate $\Delta\potentialone$ as
\begin{align}
  \Delta \potentialone \approx c \Mpl^3 \pionebgdot \Delta t
  = -c^2 \Mpl^6 \frac{f}{H} \, \Delta t\times \mathcal{O}(1) 
  = -c^2 \Mpl^6 \frac{f}{H} \, \frac{\Delta \pitwo}{M^2}\times \mathcal{O}(1) \; ,
\end{align}
where $\Delta t$ and $\Delta\pitwo$ are the time duration and the corresponding field range of the \ac{nec} violating and reheating period, $f$ is supposed to be estimated during this period, and we have assumed that the speed of $\pitwo$ takes an approximately constant value, $\pitwodot\simeq M^2$. The last assumption will be explicitly confirmed by the example in \cref{sec:A_concrete_implementation}. Let us denote by $V_2$ ($\gg \potentialone$) the effective energy density of $\pitwo$, \ie.~$V_2 \approx 3 \Mpl^2 H^2$, during the \ac{nec} violation. Then, the condition of \cref{cond_overshoot} translates to a condition on $\Delta \pitwo$, \ie.~how long the \ac{nec} violation can last in the field space, namely, neglecting order-one numerical factors,
\begin{align}
  \frac{\Delta \pitwo}{M} \lesssim \frac{M}{c^2 \Mpl} \left( \frac{\Mpl^4}{V_2} \right)^{(4m-1)/2} \frac{\Vpre}{\Mpl^4} \; . 
\end{align}
As the Universe experiences many sequences of the \ac{cc} relaxation, \ac{nec} violation and reheating, this condition must be satisfied by every sequence all the way down to the last one before the present epoch that we live today, each time with a different (decreasing) value of $\Vpre$. Barring accidental cancellation between $\Vpre$ and $\Delta\potentialone$ for the last sequence,  this requirement is fullfilled if and only if
\begin{align}
\label{eq:cond_overshoot_phi2}
  \frac{\Delta \pitwo}{M} \lesssim \frac{M}{c^2 \Mpl} \left( \frac{\Mpl^4}{V_2} \right)^{(4m-1)/2} \frac{\Lambda_{\rm obs}}{\MPlSq} \; ,
\end{align}
where $\Lambda_{\rm obs}$ is the observed value of the cosmological constant, corresponding to the present value of the $\pione$ potential $\potentialone|_{\rm now}=\MPlSq\Lambda_{\rm obs}$. 
Recalling that the parameter $m$ must take a value $m > 3/2$, this result is suggestive in the sense that, no matter how large the \ac{nec} violation is, \ie.~how large $V_2$ is, one can always achieve a sufficiently long period of it to support reheating for a sufficiently large value of $m$. We can thus pass over the overshooting problem.

%%%%%%%%%%%%%%%%%%%%%%%%%%%%%%%%%%%%%%%%%%%%%%%%%%%%%%%%%%%%%%%%%%%%%%%%%%%%
%%%%%%%%%%%%%%%%%%%%%%%%%%%%%%%%%%%%%%%%%%%%%%%%%%%%%%%%%%%%%%%%%%%%%%%%%%%%

\section{NECV and reheating sectors: background and perturbation}
\label{sec:Analysis_of_perturbative_stability}

In this section we analyze the background system of the \ac{nec}-violating and reheating sectors and the perturbations around it, in order to 
find the parameter space in which it is stable against small perturbations. We do not include the \ac{cc} relaxation sector $\pione$ in the present analysis to avoid extra computational complexity; the $\pione$ sector has supposedly decreased to a negligible amount by the time this analysis becomes relevant,\footnote{For the analysis on the perturbative stability of the $\pione$ sector alone, we would like to direct readers to \cite{Mukohyama:2003ac,Mukohyama:2003nw}.} provided that the condition \eqref{eq:cond_overshoot_phi2} is satisfied. 
Hence the action of our interest in this section is
\begin{align}
  S
  & = \int \dd^4x \sqrt{-g} \left( \mathcal{L}_{\mathrm{EH}} + \mathcal{L}_{\mathrm{NECV + reheat}} \right) \nonumber\\
  & = \int \dd^4x \sqrt{-g}
  \left[
  \frac{\Mpl^2}{2} \ScCurv
  + \KK( \pitwo , \Xtwo , \pithree )
  - \Gthree ( \pitwo , \Xtwo , \pithree ) \, \Box \pitwo
  + P ( \pithree , \Xthree )
    \right] \; ,
    \label{eq:action_pert}
\end{align}
where $\pitwo$ and $\pithree$ are scalar fields for the \ac{nec}-violating and reheating sectors, respectively.
We expand the two scalar fields as
\begin{align}
  \label{eq:decompose_fields}
  \pitwo(t , \bm{x}) & = \bar{\varphi}_2(t) + \delta\pitwo(t , \bm{x}) \; , &
  \pithree & = \bar{\varphi}_3(t) + \delta\pithree(t , \bm{x}) \; ,
\end{align}
where $\bar{\varphi}_2(t)$ and $\bar{\varphi}_3(t)$ are the homogeneous background quantities, and $\delta\pitwo$ and $\delta\pithree$ are their respective perturbations.
For the spacetime metric we conduct the ADM decomposition
\begin{equation}
  \dd s^2 = - N^2 \dd t^2 + \gamma_{ij} \left( N^i \dd t + \dd x^i \right) \left( N^j \dd t + \dd x^j \right) \; ,
\end{equation}
where $N$ and $N^i$ are the lapse and shift functions, respectively, and $\gamma_{ij}$ is the $3$-D spatial metric. Note that these quantities contain $1+3+6 = 10$ variables, which fulfills the total number of variables in the spacetime metric.
We decompose these variables into their background and perturbations as (we mainly adopt the notation in \eg.~\cite{Kobayashi:2011nu})
\begin{subequations}
  \label{eq:decompose_metric}
  \begin{align}
  N(t,\bm{x}) & = \bar{N}(t) \left[ 1 + \alpha(t,\bm{x}) \right] \; ,
  \label{eq:decompose_lapse} \\
  N^i(t,\bm{x}) & = \frac{\bar N(t)}{a(t)} \left[ \partial_i \beta(t,\bm{x}) + B^i(t,\bm{x}) \right] \; , \label{eq:decompose_shift} \\
  \begin{split}
  \gamma_{ij}(t,\bm{x}) & = a^2(t) \, \mathrm{e}^{2 \zeta(t,\bm{x})}
                          \bigg[ \delta_{ij}
                          + 2 \, \partial_i \partial_j \mathcal{E}(t,\bm{x})
                          + 2 \, \partial_{(i} E_{j)}(t, \bm{x}) \\
                          & \phantom{= a^2(t) \, \mathrm{e}^{2 \zeta(t,\bm{x})}
                          \bigg[}
                          + h_{ij}(t,\bm{x}) + \frac{1}{2} \, h_{ik}(t,\bm{x}) \, h_{kj}(t,\bm{x}) \bigg] \; ,
                          \label{eq:decompose_3Dmetric}
  \end{split}
\end{align}
\end{subequations}
where $\bar{N}$ and $a$ are the background lapse and scale factor, respectively, $\{ \alpha, \beta, \zeta, \mathcal{E}\}$ are scalar modes, the vector modes $\{ B_i , E_i\}$ satisfy the transverse conditions $\partial_i B_i = \partial_i E_i = 0$, and the tensor modes $\{ h_{ij} \}$ are transverse and traceless: $\partial_i h_{ij} = h_{ii} = 0$.
This latter condition thus brings the number of degrees of freedom of the tensor sector from $6$ to $2$.
Under the general coordinate transformation $x^\mu \to x^\mu + \xi^\mu(t, \bm{x})$, each variable transforms as
\begin{equation}
  \label{eq:gaugetrans_metric}
\begin{aligned}
  \Delta \alpha & = \frac{\partial_t ( \bar{N} \xi^0 )}{\bar{N}} \; , &
  \Delta \beta & = \frac{a^3}{\bar{N}} \, \partial_t \xi_L - a \bar{N} \xi^0 \; , &
  \Delta \zeta & = \Hubble \bar{N} \xi^0 \; , &
  \Delta \mathcal{E} & = \xi_L \; , \\
  \Delta B^i & = \frac{a}{\bar{N}} \, \partial_t \xi^i_T \; , &
  \Delta E_i & = \xi^i_T \; , &
  \Delta h_{ij} & = 0 \; ,
\end{aligned}
\end{equation}
and
\begin{align}
  \label{eq:gaugetrans_fields}
  \Delta\delta\pitwo = \partial_t\bar{\varphi}_2 \, \xi^0 \; , \qquad
  \Delta\delta\pithree = \partial_t \bar{\varphi}_3 \, \xi^0 \; ,
\end{align}
where the decomposition $\xi^i = \delta^{ij} \partial_j \xi_L + \xi^i_T$ with $\partial_i \xi^i_T = 0$ has been used, and $\Hubble \equiv \partial_t a / (a\bar{N})$ is the Hubble expansion rate.
Using the previous symmetry and appropriately choosing $\xi^\mu$, we fix the gauge by setting
\begin{align}
  \label{eq:gaugefix}
  \delta\pitwo = \mathcal{E} = E_i = 0 \; , \qquad
  \mbox{gauge fixing} \; ,
\end{align}
leaving no further gauge freedom. Then at the level of linear perturbations, the remaining vector modes $B^i$ are non-dynamical and fixed to null by constraints, and they can hence be omitted from our discussion altogether.
Thus our system of perturbations consists of the two sectors
\begin{align}
  \label{eq:variables_pert}
  \begin{cases}
    \alpha \, , \; \beta \, , \; \zeta \, , \; \delta\pithree & : \; \mbox{scalar sector} \; , \\
    h_{ij} & : \; \mbox{tensor sector} \; .
  \end{cases}
\end{align}
Among the scalar variables, $\alpha$ and $\beta$ are non-dynamical, and their values are fixed by the dynamical modes $\zeta$ and $\delta\pithree$. Therefore, each of the scalar and tensor sectors is a system of $2$ propagating degrees of freedom.
In particular, thanks to the background rotational symmetry, these sectors are decoupled at the level of the quadratic action,%
\footnote{Note that the linear action only derives the background equations and gives no information about perturbations.}
\ie.
\begin{align}
  \label{eq:action_scalartensor}
    S^{(2)} = S_{\mathrm{scalar}}[\alpha, \, \beta, \, \zeta, \, \delta\pithree] + S_{\mathrm{tensor}}[h] \; . 
\end{align}
In the following subsections, we formally derive the background equations of this two-scalar system and the stability conditions against perturbations.%
\footnote{For derivation of the equations for the background as well as perturbations, we refer to \cite{Kobayashi:2011nu} for a detailed calculations of a single-field case and to \cite{Kobayashi:2013ina} for a general multi-field extension.}

\subsection{Background equations}

The background quantities, defined in \cref{eq:decompose_fields,eq:decompose_metric}, obey their classical \ac{eom}, which are obtained by varying the action \cref{eq:action_pert} with respect to each of them. In order to avoid crowded notations, we here take $\bar{N} = 1$ and omit the bars over background quantities in this subsection. The \ac{eom}'s then read
{\allowdisplaybreaks%
\begin{subequations}\label{eq:All_EoMs}
  \begin{align}
    \begin{split}
      0 = & -\KKX \, \pitwodot^2-\PPX \, \pithreedot^2+\GthreeChi \, \pithreedot \, \pitwodot+\GthreePi \, \pitwodot^2+\KK+\PP
      \\ &
      -3 \Hubble \GthreeX \, \pitwodot^3+3 \MPlSq \Hubble^2
          \; ,
    \end{split} \label{eq:EoM_wrt_NN} \\[1em]
    \begin{split}
      0 = & -\pitwodot \left[ \GthreeChi \, \pithreedot+\pitwodot \left(\GthreeX \, \pitwodotdot+\GthreePi \right)\right] +\KK+\PP
      + \MPlSq \left( 2 \Hdot + 3 \Hubble^2\right)
        \; ,
    \end{split} \label{eq:EoM_wrt_A} \\[1em]
    \begin{split}
      0 = & -3 \, \GthreeX \left(\Hdot+\Hubble^2\right) \pitwodot^2
      -6 \Hubble^2 \GthreeX \, \pitwodot^2
      +\GthreePiX \, \pitwodot^2 \pitwodotdot+\GthreePiPi \, \pitwodot^2 \\
        & -3 \Hubble \bigg\{ \pitwodot \left[ \KKX+\pitwodotdot \left(\GthreeXX \, \pitwodot^2 + 2 \, \GthreeX \right) + \GthreePiX \, \pitwodot^2- 2 \, \GthreePi \right] \\
        & +\pithreedot \left(\GthreeXChi \, \pitwodot^2-\GthreeChi\right)\bigg \} -\KKXX \, \pitwodot^2 \pitwodotdot-\KKPiX \, \pitwodot^2-\KKX \, \pitwodotdot+\KKPi \\
        & +\pithreedot \pitwodot \left(-\KKXChi+\GthreeXChi \, \pitwodotdot + 2 \, \GthreePiChi \right) + \GthreeChiChi \, \pithreedot^2 + \GthreeChi \, \pithreedotdot \\
        & + 2 \, \GthreePi \, \pitwodotdot
        \; ,
    \end{split} \label{eq:EoM_wrt_pitwo} \\[1em]
    \begin{split}
      0 = & \, \KKChi-\pithreedotdot \left(\PPXX \, \pithreedot^2+\PPX\right)-\PPPiX \, \pithreedot^2+\PPPi \\
        & +3 \Hubble \left(\GthreeChi \, \pitwodot-\PPX \, \pithreedot\right)+\GthreeChi \, \pitwodotdot
        \; .
    \end{split} \label{eq:EoM_wrt_pithree}
  \end{align}
\end{subequations}
}%
The first \cref{eq:EoM_wrt_NN} is a constraint equation, and the remaining three equations
constitute a system of second-order differential equations for $\pitwo$, $\pithree$ and $a$
that shall be numerically solved in \cref{sec:Numerical_approach}.

We note that, in \cref{sec:A_concrete_implementation}, we restrict our interest to seeking the solution of the \ac{nec}-violating sector that gives $\Xtwo = {\rm constant}$, \ie.~$\pitwo \propto t$, in the absence of $\pithree$. This ansatz greatly simplifies the search of required forms of the functions $\KK$, $\Gthree$ and $\PP$ to achieve the target cosmic history as shown in \cref{fig:Hevol_entire} (or \cref{fig:Hevol_NECV} in the absence of $\pithree$). Before proceeding to the reconstruction procedure of those functions that is described in detail in \cref{sec:A_concrete_implementation}, we collect the conditions necessary to ensure our background solution to be stable against small perturbations in the following subsections.

\subsection{Perturbations}
\label{subsec:perturbations}

In this subsection we analyze the stability of the \ac{nec} violation and reheating sector against perturbations. As described at the beginning of this section, we decompose the perturbations of the fields and metric as in \cref{eq:decompose_fields,eq:decompose_metric}, respectively, and fix the gauge freedom as in \cref{eq:gaugefix}. At the level of linear perturbations, the scalar, vector and tensor sectors are mutually decoupled, thanks to the background rotational symmetry, and the vector sector contains no dynamical degree of freedom. We thus need only to study the decoupled scalar and tensor perturbations as in \cref{eq:action_scalartensor}, which is done separately hereafter.

\subsubsection{Tensor sector}

The tensor sector takes the standard form, that is, the action of the type shown in \cref{eq:action_pert} (without non-trivial $G_4$ or $G_5$ term in the Horndeski theory)  leads to the same form of $S_{\mathrm{tensor}}$ as the one of General Relativity, \ie.
\begin{align}
  \label{eq:Tensor_sector}
  S_{\mathrm{tensor}} & = \frac{\Mpl^2}{8} \int \bar{N} \dd t \, a^3 \dd^3x
                        \left[
                        \frac{\partial_t h_{ij} \, \partial_t h_{ij}}{\bar{N}^2}
                        - \frac{\partial_k h_{ij} \, \partial_k h_{ij}}{a^2}
                        \right] \; ,
\end{align}
where the background lapse $\bar{N}$ is put back.%
\footnote{In deriving the expression of \cref{eq:Tensor_sector}, the background equations are not used, which is the benefit of defining $h_{ij}$ by adding the quadratic term of $h_{ij}$ in \cref{eq:decompose_3Dmetric}.
Without it, the same form of $S_{\mathrm{tensor}}$ can be derived, but only after the background equations are imposed.}
The tensor sector is here trivially free from ghost and gradient instabilities. Therefore, we shall simply focus on the stability conditions in the scalar perturbations.

\subsubsection{Scalar sector}
\label{subsec:scalar}

The scalar sector consists of $4$ variables $\{\alpha, \beta, \zeta, \delta\pithree\} \equiv q$, where $\alpha$ and $\beta$ are non-dynamical and do not have their own kinetic terms, as explained around \cref{eq:variables_pert} . Fourier-transforming each variable as
\begin{align}
  q(t,\bm{x}) = \int \frac{\dd^3k}{(2\pi)^{3/2}} \, \mathrm{e}^{i \bm{k} \cdot \bm{x}} \, \hat{q}(t, \bm{k}) \; ,
\end{align}
with hat $\hat\dummycharacter$ denoting Fourier-transformed quantities,
the quadratic action $S_{\mathrm{scalar}}$ can be formally written in the form
\begin{align}
  S_{\mathrm{scalar}} & = \frac{1}{2} \int \dd t \, \dd^3k
                        \bigg[ 
                        \partial_t \hat\delta^\dagger A \, \partial_t \hat\delta
                        + \left( \partial_t \hat\delta^\dagger B \, \hat\delta + \mathrm{h.c.} \right)
                        + \hat\delta^\dagger C \, \hat\delta
  \nonumber\\
                      & \qquad\qquad\qquad\quad
                        + \hat{\mathcal{N}}^\dagger D \hat{\mathcal{N}}
                        +
                        \left( \hat{\mathcal{N}}^\dagger E \, \partial_t \hat\delta  + \mathrm{h.c.} \right)
                        + \left( \hat{\mathcal{N}}^\dagger F \, \hat\delta + \mathrm{h.c.} \right)
                        \bigg] \; ,
                        \label{eq:action_scalar}
\end{align}
up to total derivatives,
where $\hat\delta = \{ \hat\zeta, \delta\hat\pithree \}$ and $\hat{\mathcal{N}} = \{ \hat\alpha, \hat\beta \}$ are arrays grouping the dynamical and non-dynamical variables, respectively.
Note that the reality condition of the variables in the coordinate space is translated to $\hat{q}^\dagger (\bm{k}) = \hat{q} (- \bm{k})$ in the Fourier space.
The coefficients $A, B, C, D, E$ and $F$ are all $2\times 2$ square matrices whose components consist of the background quantities, explicitly
{\allowdisplaybreaks%
\begin{align}
    \begin{aligned}
        A & = \frac{a^3}{N}
            \begin{pmatrix}
                - 6 \Mpl^2
                & 0 \\
                0
                & \mathcal{G}_\varphi
            \end{pmatrix} \; , &
        B & = a^3
            \begin{pmatrix}
                0
                & \displaystyle \frac{3}{2} \, c_1 \\
                - \displaystyle \frac{3}{2} \, c_1
                & 0
            \end{pmatrix} \; , \\
        C & = N a^3
            \begin{pmatrix}
                \displaystyle
                2 \Mpl^2 \, \frac{k^2}{a^2}
                & \displaystyle
                \frac{3}{2} \, c_2 \vspace{1mm}\\
                \displaystyle
                \frac{3}{2} \, c_2
                & \displaystyle
                - \frac{k^2}{a^2} \left( \mathcal{G}_\varphi - 2 \Xthree \, \PPXX \right) + c_3
            \end{pmatrix} \; , &
        D & = N a^3
            \begin{pmatrix}
                \displaystyle
                2 \Sigma  + \frac{c_4^2}{\mathcal{G}_\varphi}
                & \displaystyle
                \frac{2 k^2}{a} \, \Theta \vspace{1mm} \\
                \displaystyle
                \frac{2 k^2}{a} \, \Theta
                & 0
            \end{pmatrix} \; , \\
        E & = a^3
            \begin{pmatrix}
                6 \Theta
                & c_4 \vspace{1mm} \\
                \displaystyle
                - 2 \Mpl^2 \, \frac{k^2}{a}
                & 0
            \end{pmatrix} \; , &
        F & = N a^3
            \begin{pmatrix}
                \displaystyle
                2 \Mpl^2 \, \frac{k^2}{a^2}
                & c_5  \\
                0
                & \displaystyle
                \frac{k^2}{a} \, c_1
            \end{pmatrix} \; ,
    \end{aligned}
\end{align}
}%
where
{\allowdisplaybreaks%
\begin{align}
  \mathcal{G}_{\varphi} & = \PPX + 2 \Xthree \, \PPXX  \; , \label{eq:Gvarphi}\\
  \begin{split}
    \Sigma & = -3 \Mpl^2 \Hubble^2 - \frac{\Xtwo \left( \GthreeChi \right)^2}{\mathcal{G}_{\varphi}}
             + \Xtwo \bigg( \KKX + 2 \Xtwo \, \KKXX - 2 \, \GthreePi \\
             & \quad + 12 \Hubble \, \frac{\pitwodot}{N} \, \GthreeX - 2 \Xtwo \, \GthreePiX + 6 \Hubble \, \frac{\pitwodot}{N} \, \Xtwo \, \GthreeXX
                    - \frac{\pitwodot \pithreedot}{N^2} \, \GthreeXChi \bigg) \; ,
  \end{split} \\
  \begin{split}
    \Theta & = \Mpl^2 \Hubble - \frac{\pitwodot}{N} \, \Xtwo \, \GthreeX
  \end{split} \\
    c_1 & = \frac{\pitwodot}{\bar{N}} \, \GthreeChi - \frac{\pithreedot}{\bar{N}} \, \PPX \; , \\
  \begin{split}
    c_2 & = - \left[ \frac{1}{N} \, \partial_t \left( \frac{\pitwodot}{N} \right) + 3 \Hubble \, \frac{\pitwodot}{N} \right] \GthreeChi
          - \frac{2}{N} \, \partial_t \left( \frac{\pitwodot}{N} \right) \Xtwo \GthreeXChi
          - 2 \Xtwo \GthreePiChi \\
          & \quad
            - \frac{\pitwodot \pithreedot}{N^2} \, \GthreeChiChi
            + \left[ \frac{1}{N} \, \partial_t \left( \frac{\pithreedot}{N} \right) + 3 \Hubble \, \frac{\pithreedot}{N} \right] \PPX
          + \frac{2}{N} \, \partial_t \left( \frac{\pithreedot}{N} \right)  \Xthree \PPXX \\
          & \quad + 2 \Xthree \PPPiX \; ,
  \end{split} \\
  \begin{split}
    c_3 & = \KKChiChi
          + \left[ \frac{1}{N} \, \partial_t \left( \frac{\pitwodot}{N} \right) + 3 \Hubble \, \frac{\pitwodot}{N} \right] \GthreeChiChi
          + \PPPiPi
          - 2 \Xthree \PPPiPiX \\
          & \quad - \left[ \frac{1}{N} \, \partial_t \left( \frac{\pithreedot}{N} \right) + 3 \Hubble \, \frac{\pithreedot}{N} \right] \PPPiX
          - \frac{2}{N} \, \partial_t \left( \frac{\pithreedot}{N} \right) \Xthree \PPPiXX \; ,
  \end{split} \\
  c_4 & = \frac{\pitwodot}{N} \, \GthreeChi - \frac{\pithreedot}{N} \, \mathcal{G}_{\varphi} \; , \\
  \begin{split}
    c_5 & = -2 \Xtwo \KKXChi
          - \left[ \frac{1}{N} \, \partial_t \left( \frac{\pitwodot}{N} \right) + 3 \Hubble \, \frac{\pitwodot}{N} \right] \GthreeChi \\
          & \quad - 6 \Hubble \, \frac{\dot\varphi_2}{N} \, \Xtwo \GthreeXChi
          + 2 \Xtwo \GthreePiChi
          + \frac{\pitwodot \pithreedot}{N^2} \, \GthreeChiChi \\
          & \quad + \left[ \frac{1}{N} \, \partial_t \left( \frac{\pithreedot}{N} \right) + 3 \Hubble \, \frac{\pithreedot}{N} \right] \PPX
          + \frac{2}{N} \, \partial_t \left( \frac{\pithreedot}{N} \right)  \Xthree \PPXX \; ,
  \end{split}
\end{align}
}%
noting that the bar $\bar\dummycharacter$ is here omitted from the background quantities to avoid crowded notation.

By varying the action of \cref{eq:action_scalar} with respect to $\hat{\mathcal{N}}^\dagger$, we find the constraint equations that fix the non-dynamical variables in terms of the dynamical ones $\delta$, given by
\begin{align}
  \hat{\mathcal{N}} = - D^{-1} \left( E \, \partial_t \hat\delta + F \hat\delta \right) \; ,
\end{align}
as $\det(D) \ne 0$ in our current system.
Plugging this back into \cref{eq:action_scalar}, we obtain the action in terms only of the dynamical degrees of freedom,
\begin{align}
  S_{\mathrm{scalar}} & = \frac{1}{2} \int \dd t \, \dd^3k
                        \left[
                        \partial_t \hat\delta^\dagger \tilde{A} \, \partial_t \hat\delta
                        +
                        \left( \partial_t \hat\delta^\dagger \tilde{B} \, \hat\delta + \mathrm{h.c.} \right)
                        + \hat\delta^\dagger \tilde{C} \, \hat\delta
                        \right] \; ,
\end{align}
where
\begin{align}
  \tilde{A} = A - E^\dagger D^{-1} E \; , \qquad
  \tilde{B} = B - E^\dagger D^{-1} F \; , \qquad
  \tilde{C} = C - F^\dagger D^{-1} F \; .
\end{align}
Notice that now the kinetic matrix $\tilde{A}$ is no longer diagonal. In order to diagonalize it, we can perform the following change of variables,
\begin{align}
  \hat\delta = R \hat\Delta \; , \qquad
  R =
  \left(
  \begin{array}{cc}
    1 & 0 \vspace{1mm} \\
    \displaystyle
    - \frac{\Mpl^2 c_4}{\mathcal{G}_{\varphi} \Theta} & 1
  \end{array}
  \right) \; ,
\end{align}
which results in the action in terms of $\hat\Delta$,
\begin{align}
  S_{\mathrm{scalar}} & = \frac{1}{2} \int \dd t \, \dd^3k
                        \left[
                        \partial_t \hat\Delta^\dagger \tilde{T} \, \partial_t \hat\Delta
                        +
                        \left( \partial_t \hat\Delta^\dagger \tilde{X} \, \hat\Delta + \mathrm{h.c.} \right)
                        - \hat\Delta^\dagger \tilde\Omega^2 \, \hat\Delta
                        \right] \; ,
\end{align}
where
\begin{align}
    \begin{aligned}
        \tilde T & = R^\dagger \tilde{A} R \; , \\
        \tilde X & = R^\dagger \tilde{B} R + R^\dagger \tilde{A} \, \partial_t R \; , \\
        \tilde\Omega^2 & = - R^\dagger \tilde{C} R - \partial_t R^\dagger \tilde{A} \, \partial_t R - \partial_t R^\dagger \tilde{B} R - R^\dagger \tilde{B}^\dagger \partial_t R \; .
    \end{aligned}
\end{align}
Finally, by adding the total derivative $- \frac{1}{4} \int \dd t \, \dd^3k \left( \tilde{X} + \tilde{X}^\dagger \right)$,
we arrive at the final expression for the quadratic action
\begin{align}
  \label{eq:Scalar_perturbation_Lagrangian_after_Diagonalisation}
  S_{\mathrm{scalar}} & = \frac{1}{2} \int \dd t \, \dd^3k
                        \left[
                        \partial_t \hat\Delta^\dagger T \, \partial_t \hat\Delta
                        + \partial_t \hat\Delta^\dagger X \, \hat\Delta
                        - \hat\Delta^\dagger X \, \partial_t \hat\Delta
                        - \hat\Delta^\dagger \Omega^2 \, \hat\Delta
                        \right] \; ,
\end{align}
where
\begin{align}
  T = \tilde{T} \; , \qquad
  X = \frac{\tilde{X} - \tilde{X}^\dagger}{2} \; , \qquad
  \Omega^2 = \tilde\Omega^2 + \frac{\partial_t \tilde{X} + \partial_t \tilde{X}^\dagger}{2} \; .
\end{align}
Note that $T$ and $\Omega^2$ are symmetric $2\times 2$ matrices, and $X$ is an anti-symmetric one.
Now the kinetic matrix
\begin{align}
  \label{eq:matrix_T}
  T & = \frac{a^3}{N}
  \left(
  \begin{array}{cc}
    \displaystyle
    2 \Mpl^2 \, \mathcal{G}_S & 0 \vspace{1mm} \\
    0 & \displaystyle \mathcal{G}_{\varphi}
  \end{array}
  \right) \; , \qquad
  \mathcal{G}_S \equiv 3 + \Mpl^2 \, \frac{\Sigma}{\Theta^2} \; ,
\end{align}
is diagonal, as desired.
The conditions to prohibit ghost instabilities are therefore
\begin{align}
  \label{eq:noghost}
  \mathcal{G}_S & > 0 \; , \qquad
  \mathcal{G}_{\varphi} > 0 \; , \qquad\quad
  \text{no ghost} \; ,
\end{align}
serving as part of the conditions of perturbative stability ($\mathcal{G}_{\varphi}$ is defined in \cref{eq:Gvarphi}).

Another instability we need to suppress is the gradient instability. We are only concerned with high-momentum catastrophic instability, as the low-momentum counterpart may be harmless at a non-linear level \textit{\`a la} Jeans instability.
This amounts to imposing positivity condition for the squared sound speeds of the scalar perturbations, $\cstwo$, as defined below.
In the high-$k$ limit, we collect the terms in the coefficient matrices of the order $T \sim \bigO(k^0)$, $X \sim \bigO(k^1)$ and $\Omega^2 \sim \bigO(k^2)$. The full expression of $T$ is given in \cref{eq:matrix_T} and does not depend on $k$. The other matrices in the high-$k$ limit read
\begin{align}
  X & = 0 + \bigO(k^0) \; , \qquad
  \Omega^2 = \Omega^2_{\bigO(k^2)} + \bigO(k^0) \; ,
\end{align}
where $\Omega^2_{\bigO(k^2)}$ only contains the terms proportional to $k^2$ in the high-$k$ limit. Taking the ansatz $\hat\Delta \propto \exp\left( i \int^t N \dd t' \soundspeed  k / a \right)$ as an adiabatic solution for large $k$, the sound speed $\cstwo$ can be found by solving the characteristic equation
\begin{align}
  \det \left(- \cstwo k^2 \, \frac{N^2}{a^2} \,  T + \Omega^2 \right) = 0 \; ,
\end{align}
leading to
\begin{align}
  \label{eq:cs2}
  \soundspeed^4
  - \left( c_\chi^2 + \frac{\mathcal{F}_S}{\mathcal{G}_S} + \mathcal{A} + \mathcal{B} \right) \cstwo
  + c_\chi^2 \left( \frac{\mathcal{F}_S}{\mathcal{G}_S} + \mathcal{A} \right)
  = 0 \; ,
\end{align}
where
\begin{align}
    \begin{aligned}
    \mathcal{F}_S & \equiv \frac{\Mpl^4}{N a} \, \partial_t \left( \frac{a}{\Theta} \right) - \Mpl^2
    \; , & c_\chi^2 & \equiv \frac{\PPX}{\PPX + 2 \Xthree \, \PPXX} \; , \\
    \mathcal{A} & \equiv - \frac{\Mpl^4}{2 \PPX \, \mathcal{G}_S \, \Theta^2} \left( \frac{\pitwodot}{N} \, \GthreeChi - \frac{\pithreedot}{N} \, \PPX  \right)^2
    \; , & \mathcal{B} & \equiv \frac{4 \Mpl^4 \Xtwo \Xthree^2 \left( \PPXX \right)^2 \left( \GthreeChi \right)^2}{\PPX \, \mathcal{G}_S \, \mathcal{G}_{\varphi}^2 \Theta^2} \; .
    \end{aligned}
\end{align}
The values of $\cstwo$ are determined by the two roots of \cref{eq:cs2}, and we impose the conditions
\begin{align}
  \label{eq:nograd}
  \cstwo > 0 \; , \qquad
  \mbox{gradient stability} \; ,
\end{align}
for each value of $\cstwo$.

Notice from \cref{eq:cs2} that, in the particular cases of $\GthreeChi = 0$ and/or $\PPXX = 0$, we have $\mathcal{B} = 0$, and the expressions of $\cstwo$ vastly simplify. In our numerical examples, we indeed take
\begin{align}
  \GthreeChi = 0 \qquad \& \qquad
  \PPXX = 0 \; , \qquad
  \mbox{ numerical examples} \; ,
\end{align}
and in this case
\begin{align}
\label{eq:cs2_simple}
  \cstwo = c_\chi^2 \text{\quad or \quad} \cstwo = \frac{\mathcal{F}_S}{\mathcal{G}_S} + \mathcal{A} \; ,
\end{align}
while
\begin{align}
  c_\chi^2 = 1 \; , \qquad
  \mathcal{A} = - \frac{\Mpl^4 \Xthree \, \PPX}{\mathcal{G}_S \, \Theta^2} \; .
\end{align}
Therefore, in these particular examples, we only need to impose
\begin{align}
  \label{eq:nograd_simple}
  \mathcal{F}_S > \frac{\MPl^4 \Xthree \PPX}{\Theta^2} \; ,
\end{align}
to avoid gradient instability, as we have already imposed $\mathcal{G}_S > 0$ from the no-ghost condition.
In summary, to ensure the background evolution to be stable against small perturbations, we impose the conditions in \cref{eq:noghost,eq:nograd}, or those in  \cref{eq:noghost,eq:nograd_simple} for the particular and simpler examples, for the entire history of the Universe, and particularly during the phases of \ac{nec} violation and reheating.

%%%%%%%%%%%%%%%%%%%%%%%%%%%%%%%%%%%%%%%%%%%%%%%%%%%%%%%%%%%%%%%%%%%%%%%%%%%%
%%%%%%%%%%%%%%%%%%%%%%%%%%%%%%%%%%%%%%%%%%%%%%%%%%%%%%%%%%%%%%%%%%%%%%%%%%%%

\section{A concrete implementation}
\label{sec:A_concrete_implementation}

We have until now kept arbitrary the free functions $f_{1,2}(\pitwo,\pithree)$, $f(\pitwo,\pithree)$, $V(\pitwo,\pithree)$ and $U(\pithree)$ in the \ac{nec} violating and reheating Lagrangian $\LNECVreh$, \cref{eq:NECVreheat}, but we now choose a concrete set of functions to fully realise our desired cosmological scenario.
For the purpose of the conceptual proof, we take an inverted route: we first determine a desired cosmic history we would like to achieve, which is summarized in \cref{sec:Overall_picture_of_the_cosmic_history}, and then reconstruct the functions accordingly. While we disregard the reheating field $\varphi_3$ in this procedure, we numerically verify the overall behavior of the whole system after reviving the $\varphi_3$ dependence into the reconstructed model.

\subsection{Reconstruction of the NEC-violating sector}
\label{subsec:reconst_necv}

We have a total of $3$ independent functions, $K$, $G_3$ and $P$, and expand $K$ and $G_3$ as polynomial functions of $\Xtwo$ as in \cref{eq:KK_and_G3_with_pithree_turned_off}.
To reconstruct them, we first consider the \ac{nec}-violating sector alone by turning off the $\varphi_3$ and disregarding $P(\varphi_3, \Xthree)$.
On the one hand, there are $4$ independent functions, $\tilde{f}_1(\pitwo)$, $\tilde{f}_2(\pitwo)$ and $\tilde{V}(\pitwo)$ from $\tilde{K}$, and $\tilde{f}(\pitwo)$ from $\tilde{G_3}$ (recall tilde denotes quantities with $\varphi_3$ dependence taken out). On the other hand, we have $2$ independent background equations from \cref{eq:All_EoMs} (with $\varphi_3$ turned off). Therefore, the forms of two of these functions can be fixed by the background equations, once all other arbitrary functions are chosen by hand. In the following we reconstruct these functions in the action so that the system admits the following solution,
\begin{align}
  \label{eq:phi2_solution}
 H = H_{\rm necv}(\pitwo)\;, \quad  \pitwo = t \; , \quad \bar{N} = 1\;, 
\end{align}
where $H$ is the Hubble expansion rate and $H_{\rm necv}(\pitwo)$ is a fixed function corresponding to the input Hubble expansion rate.

The \ac{nec}-violating sector is assumed to be periodic in the field value of $\pitwo$, as discussed in \cref{subsec:NECreheat}. Without loss of generality, we shift $\pitwo$ such that the \ac{nec}-violating (and reheating) period is localized around $\pitwo = 0$. We then take the following ansatz for the forms of $\tilde{K}$ and $G_3$, that is,
\begin{subequations}
  \label{eq:tilde_KG3}
\begin{align}
  \tilde{K} & = \Mpl^2 H_{\rm dip}^2 \left[ F_1(\pitwo) \, \Xtwo + F_2(\pitwo) \, \Xtwo^2 - v(\pitwo) \right] \; , \quad
              v(\pitwo) = -
              v_0 \exp \left( - \frac{\pitwo^2}{2 T_{\rm dip}^2} \right) \; ,
              \label{eq:Ktilde} \\
  \tilde{G}_3 & = \Mpl^2 H_{\rm necv}(\pitwo) F_{\rm kb}(\pitwo) \, \Xtwo \; , \qquad
                F_{\rm kb}(\pitwo) = F_{\rm kb,0} + F_{\rm kb,1} \exp \left( - \frac{\pitwo^2}{2 T_{\rm kb}^2} \right) \; ,
                \label{eq:G3tilde}
\end{align}
\end{subequations}
where $\Hdip$ is a constant of mass dimension $1$, $\Tdip$ and $\Tkb$ of mass dimension $-1$, and $v_0$, $F_{\rm kb,0}$  and $F_{\rm kb, 1}$ are dimensionless constants. We here normalize $\pitwo$ so that it has mass dimension $-1$, and thus $\Xtwo$ is dimensionless. The functions $F_1$, $F_2$ and $\Fkb$ correspond to $\tilde{f}_1$, $\tilde{f}_2$ and $\tilde{f}$ (\cref{eq:KK_and_G3_with_pithree_turned_off}), respectively, that are made dimensionless. The potential $v$ has a \enquote{dip} of depth $v_0$ at $\pitwo = 0$ so that $\pitwo$ can be trapped to sustain a \ac{nec}-violating period. The $\tilde{G}_3$ contribution is needed to ensure the stability of the system in this period. For this reason, we use the input Hubble expansion rate $H_{\rm necv}(\pitwo)$ as an overall factor of $\tilde{G}_3$ so that the $\tilde{G}_3$ contribution becomes prominent in the \ac{nec} violating phase.

%%%%%%%%%%%%%%%%%%%%%%%%%%%
%%%%%%%%%%%%%%%%%%%%%%%%%%%
\begin{figure}
    \centering
    \tikzsetnextfilename{Hguess}
\begin{tikzpicture}[]
    \begin{axis}[
        title={Input Hubble parameter $\Hinput$},
        width={0.7\textwidth},
        height={0.3\textheight},
        xlabel={$10^{-4} \cdot \pitwo$},
        xmajorgrids={true},
        xmin=-2.5,
        xmax=2.5,
        ylabel={},
        ymajorgrids={true},
        ymin={-0.025},
        ymax={1},
    ]
        \addplot[
            color=colorhubble,
            % opacity=0.7,
            % dashed,
            % thin,
            thick,
        ] table[col sep=comma,header=false, x expr=(\thisrowno{0}-2.5e4)/1e4, y index=1] {CSVs/Hguess.csv};
    \end{axis}
\end{tikzpicture}
    \caption{
      Overall shape of the input Hubble expansion rate $H_{\rm necv}$.
      The plot uses the numerical parameters exhibited in \cref{tab:Numerical_parameters}.
    }
    \label{fig:Hinput_overall_shape}
\end{figure}
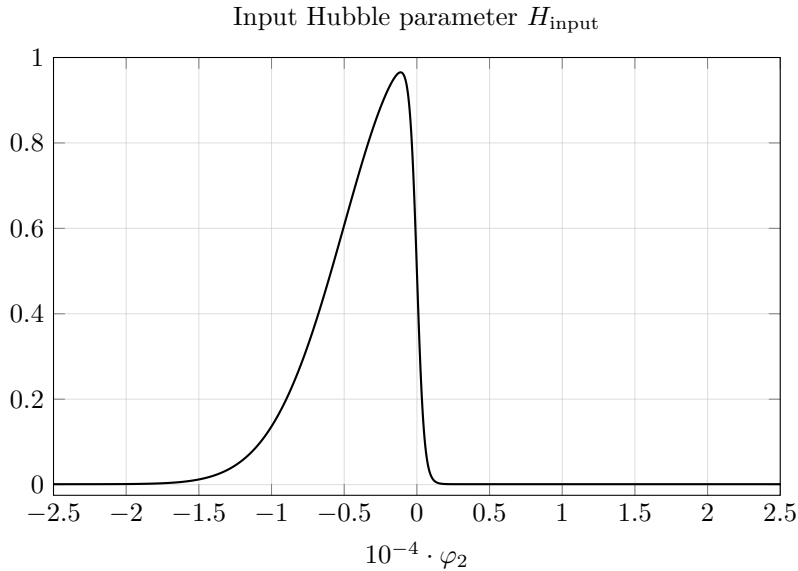
%%%%%%%%%%%%%%%%%%%%%%%%%%%
%%%%%%%%%%%%%%%%%%%%%%%%%%%
%
For the actual form of $H_{\rm necv}$ that we employ, we introduce two scales to turn on and off \ac{nec} violation. There are many possible forms to achieve this, and we simply take one particular choice, \ie.
\begin{align}
  \label{eq:Hnecv}
  H_{\rm necv}(\pitwo) = H_0 + H_1 \exp\left( - \frac{\pitwo^2}{2 T^2} \right) \frac{1 - \tanh \left( \frac{\pitwo}{\tau} \right)}{2} \; ,
\end{align}
where $H_0$ and $H_1$ are constants of dimension $1$, while $T$ and $\tau$ those of dimension $-1$. The value of $H_{\rm necv}$ far away from the origin (\ie.~$\vert \pitwo \vert \gg T, \tau$) is $H_0$, that is during a major duration of the cosmic history, and is essentially the present value of the Hubble expansion rate if we apply this system to the last \ac{nec} violating phase before the present epoch of the Universe. However, during the \ac{nec} violating phase, it ascends to the maximum value $\sim H_0 + H_1$ within a time scale controlled by $T$, eventually going through a step-function-like drop around $\pitwo = 0$ to $H_0$ with another scale controlled by $\tau$.
The overall shape of $H_{\rm necv}$ is graphically depicted in \cref{fig:Hinput_overall_shape}.
In practice, the drop needs to be sharper than the rise ($T \gg \tau$) in order to accommodate a reheating period toward the end of the \ac{nec} violation by transferring a large portion of the energy in $\pitwo$ abruptly enough so that the reheating field $\varphi_3$ starts oscillating using the transferred energy.

We now have all the setups for the reconstruction. In order to ensure the target cosmic history of \cref{eq:Hnecv} to be compatible with time evolution, we seek for a solution of the form \cref{eq:phi2_solution}. We then substitute \cref{eq:phi2_solution} and \cref{eq:tilde_KG3} into the background equations of motion in \cref{eq:All_EoMs}. Only two of the equations are independent when $\varphi_3$ is turned off, and they give
\begin{subequations}
  \label{eq:F1F2}
\begin{align}
  F_1 & = 4 v + 3 \left( F_{\rm kb} - 4 \right) \frac{H_{\rm necv}^2}{H_{\rm dip}^2} + \frac{F_{\rm kb}' H_{\rm necv}}{H_{\rm dip}^2} + \left( F_{\rm kb} - 6 \right) \frac{H_{\rm necv}'}{H_{\rm dip}^2} \; , \\
  F_2 & = - 4 v - 6 \left( F_{\rm kb} - 2 \right) \frac{H_{\rm necv}^2}{H_{\rm dip}^2} + \frac{4 H_{\rm necv}'}{H_{\rm dip}^2} \; ,
\end{align}
\end{subequations}
where prime denotes $\partial / \partial t = \partial / \partial \pitwo$ under \cref{eq:phi2_solution}. Recovering $\pitwo$ dependence by basically replacing $t \to \pitwo$, these two equations fix the forms of $F_1$ and $F_2$, given the predetermined functions $H_{\rm necv}$, $F_{\rm kb}$ and $v$.
This completes the reconstruction of the \ac{nec}-violating sector for our concrete implementation of the model.

\subsection{Attractor behavior of the NEC-violating dynamics}
\label{sec:Stability_and_attractor}

We have fixed the forms of the model functions in the $\pitwo$ sector as \cref{eq:F1F2} so that the \ac{nec} violation as depicted in \cref{eq:Hnecv} and \cref{eq:phi2_solution} is achieved as a solution of the background equations of motion. However whether or not this particular solution is an attractor of the dynamical system is a separate issue, which we would like to address in this subsection. The analysis here concerns the background \cref{eq:All_EoMs} with the reconstructed functions of \cref{eq:tilde_KG3} with \cref{eq:Hnecv,eq:F1F2}.

We first perform a linearized analysis, and to this end we expand the background quantities as
\begin{align}
  \label{eq:background_perturb}
  \pitwo(t) = \varphi^{(0)}(t) + \epsilon \, \varphi^{(1)}(t) \; , \qquad
  H(t) = H^{(0)}(t) + \epsilon \, H^{(1)}(t) \; ,
\end{align}
where $\varphi^{(0)} = t$ and $H^{(0)} = H_{\rm necv}(t)$ denote the solutions assumed for the reconstruction in the previous subsection while $\epsilon\varphi^{(1)}$ and $\epsilon H^{(1)}$ are small perturbations, and $\epsilon$ is the expansion parameter. We then expand the background \cref{eq:All_EoMs} up to the linear order in $\epsilon$. The $\mathcal{O}(\epsilon^0)$ equations are trivially satisfied, and the $\mathcal{O}(\epsilon^1)$ equation for $H^{(1)}$ is a constraint equation that can be solved for $H^{(1)}$ in favor of $\varphi^{(1)}(t)$ and $\partial_t \varphi^{(1)}$. As a result, the master equation of the linearized system is only in terms of $\varphi^{(1)}$ with $\mathcal{O}(\epsilon^0)$ coefficients, and reducing it to a coupled system of first-order equations gives, in a matrix form,
\begin{align}
  \label{eq:background_linearequation}
  \partial_t \left(
  \begin{array}{c}
    \varphi^{(1)} \\ \pi_\varphi^{(1)}
  \end{array}
  \right)
  =
  \mathcal{M}
  \left(
  \begin{array}{c}
    \varphi^{(1)} \\ \pi_\varphi^{(1)}
  \end{array}
  \right) \; , \qquad
  \mathcal{M} = \left(
  \begin{array}{cc}
    0 & 1 \\
    \mathcal{A} & \mathcal{B}
  \end{array}
  \right) \; ,
\end{align}
where $\mathcal{A}$ and $\mathcal{B}$ are functions of the $\mathcal{O}(\epsilon^0)$ quantities.
While $\mathcal{A}$ and $\mathcal{B}$ are in general of lengthy expressions, our restriction of including only up to the $\mathcal{L}_3$ terms of Horndeski theory with the \ac{eom}-consistent reconstructed functions of \cref{eq:tilde_KG3} conveniently sets $\mathcal{A} = 0$. Hence, the eigenvalues of the matrix $\mathcal{M}$ are $0$ and $\mathcal{B}$.
In order for the solutions $\varphi^{(0)} = t$ and $H^{(0)} = H_{\rm necv}(t)$ to be a local attractor of the system, we thus require
these eigenvalues be non-positive, \ie.%
\footnote{In principle the non-positiveness condition could be imposed only on the real part of the eigenvalues. However, the components of $\mathcal{B}$ (and $\mathcal{A}$) consist of the $\mathcal{O}(\epsilon^0)$ quantities, and the only occasion in which they become complex is when the value of $H^{(0)}$ becomes complex as a solution of the constraint equation, but this only means that there is no real solution at the $\mathcal{O}(\epsilon^0)$ order. We exclude such a case.}
\begin{align}
  \label{eq:cond_attractor}
  \mathcal{B} \le 0 \; .
\end{align}
This ensures the local stability of the solution assumed in the previous subsection.

\begin{figure}
    \centering
    \tikzsetnextfilename{phaseportrait}
\begin{tikzpicture}
    \begin{axis}[
        width={0.7\textwidth},
        height={0.3\textheight},
        xlabel={$10^{-4} \cdot \pitwo$},% \cdot 10^{-4}
        xmajorgrids={true},
        xmin=-2.5,
        xmax=2.5,
        ylabel={$\pitwodot$},
        ymajorgrids={true},
        ymin={0.95},
        ymax={1.05},
        axis on top=true,
    ]
        \addplot[
            fill=LightGray,
            postaction={
                pattern=dots, %north east lines
                pattern color=Black,
            }
        ] table[col sep=comma,header=false,x expr=\thisrowno{0}/1e4,y index=1] {CSVs/imaginary_region1.csv};
        \draw[densely dotted, thin] (axis cs:-0.1,1) -- (axis cs:0.1,1);
        \foreach \i in {1,2,3,4,...,997}{
            \addplot[black,-stealth=0.3,very thin] table[
                col sep=comma,
                header=false,
                x expr=\thisrowno{0}/1e4,
                y index=1,
            ]{CSVs/phaseportrait-lines/line\i.csv};x
        }
        \draw[thin] ({axis cs:0,1}-|{rel axis cs:0,0}) -- ({axis cs:0,1}-|{rel axis cs:1,0});
    \end{axis}
\end{tikzpicture}
    \caption{
      The phase portrait of $\pitwo$ and $\pitwodot$ as a vector field.
      Arrows are scaled proportionally to the gradient intensity.
      The above plot therefore exhibits the strong attraction to the solution $\pitwo \equiv t$, especially in the central area (between approximately $\pitwo=\num{-5000}$ and $\num{0}$).
      The lightly shaded regions are where $H$ takes a complex value as a solution to the constraint \cref{eq:EoM_wrt_NN}, that is where a real background solution does not exist.
    }
    \label{fig:Phase_portrait}
\end{figure}
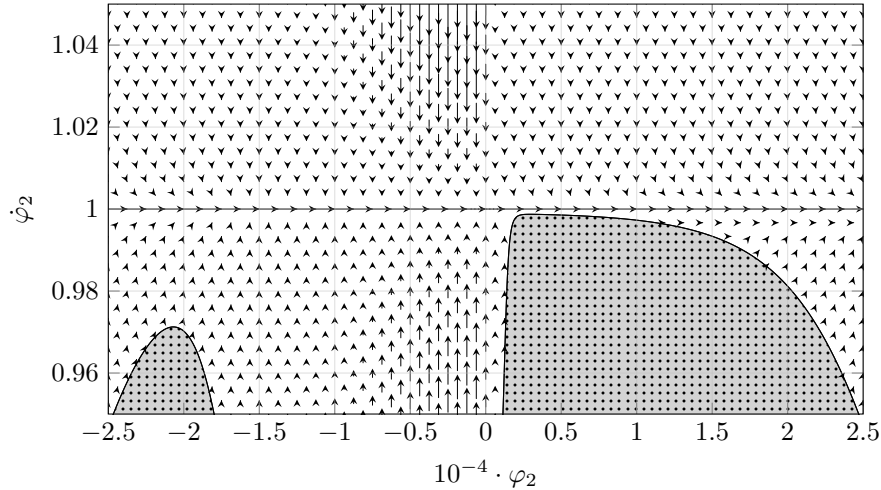
The previous analysis is a linearized one, and in order to observe a more global behavior of the system, we resort to a numerical phase-space illustration, a representative case of which is depicted in \cref{fig:Phase_portrait}. As seen in the figure, $\pitwo = t$, which is our input solution for the reconstruction, is indeed the global attractor of the system. This justifies our ansatz and reconstruction of the model functions in the previous subsection, and the system is resistant against small perturbations at least in the direction of positive $\dot{\varphi}_2$.
In \cref{fig:Phase_portrait}, the values of $H_0$, which is essentially the Hubble expansion rate at present, is taken rather close to that of $H_1$, which corresponds to the amount of raise in $H$ during the \ac{nec} violating phase, for numerical ease. When the hierarchy between these values is closer to a realistic one, we note two cautions. First, the shaded complex regions approach closer to the $\dot{\varphi}_2$ line. However they never cross it provided taking sufficiently large value of $v_0$. Second, when $H_0$ becomes much smaller than $H_1$, the attractor behavior towards the input solution $\pitwo=t$ is rather weak away from the NEC-violating region.
Indeed, in a small $H_0$ limit and sufficiently far from the NEC violation so that $v \approx 0$ and $F_{\rm kb} \approx F_{\rm kb,0}$, the value of $\mathcal{B}$ takes
\begin{align}
  \label{eq:2}
  \mathcal{B} \simeq - 3 H_0 \; , \qquad
  H_0 \ll H_1 \; ,
\end{align}
which makes the validity of the solution $\pitwo = t$ somewhat worrisome. However, it is still a local attractor because $\mathcal{B} < 0$. Moreover, we are assuming a period of the \ac{nec}-preserving phase much longer than that of the \ac{nec}-violating one. Therefore, even though the attractor appears to be weak in the limit $H_0 \ll H_1$, the background system is expected to go back to the solution $\pitwo = t$ during the long \ac{nec}-preserving era.

\subsection{Recovering the reheating sector coupled to NEC violation}
\label{subsec:recover_reheating}

To realize the known big bang universe after the \ac{nec}-violating era, the reheating field $\varphi_3$ necessarily couples to $\pitwo$ in a non-minimal manner. Our choice of the coupling amounts to making the following replacements: 
\begin{subequations}
    \label{eq:coupling_phi23}
\begin{align}
  F_1 & \to \frac{1 + \alpha_{\rm kick} \, {\rm e}^{- \beta_{\rm kick} \varphi_3}}{1 + \alpha_{\rm kick}} \, F_1 \; , \\
  F_2 & \to \frac{1 + \alpha_{\rm kick} \, {\rm e}^{- \beta_{\rm kick} \varphi_3}}{1 + \alpha_{\rm kick}} \, F_2\; , \\
  v & \to \exp\left[ - \beta_{\rm dip}^2 \left( \left( \varphi_3 - 1 \right)^2 - 1 \right) \right] v \; ,
\end{align}
\end{subequations}
while we reintroduce the kinetic term and bare potential of $\varphi_3$ as
\begin{align}
  \label{eq:form_P}
  P(\varphi_3, \Xthree) = \Mpl^2 \beta_{\rm kin}^2 \, \Xthree - U(\varphi_3) \; , \qquad
  U(\varphi_3) = 3 \Mpl^2 H_I^2 \left( 1 - {\rm e}^{-\betaI \varphi_3} \right)^2 \; ,
\end{align}
where $\varphi_3$ is normalized to be dimensionless, $\alphakick$, $\betakick$, $\betadip$, $\betakin$ and $\betaI$ are all dimensionless parameters, and $H_I$ is a constant of mass dimension $1$. Note that setting $\varphi_3 = 0$ gives the case of the \ac{nec}-violating sector alone studied in the previous subsection.

The explicit but rather nontrivial forms of the interactions above are taken to accomplish the desired efficient energy transfer from the \ac{nec}-violating sector to reheat the Universe.
One feature we aim for is to eventually yield to an oscillating reheating field $\pithree$ (as explained with \cref{fig:Reheating_potential_schematic}, and numerically verified in \cref{fig:Numerical_chi_and_chi_dot}).
The above choices lead to the following dynamics:
\begin{itemize}
\item When the \ac{nec} is well preserved, $\varphi_3$ is effectively decoupled from $\pitwo$. Then, $\varphi_3$ stays at the minimum of its Starobinsky-type bare potential $U$, \ie.~$\varphi_3 = 0$ while the \ac{nec} holds.
\item During the time when the \ac{nec} violation takes place, the different potentials $\couplingpotential$ and $\potentialthree$ together shape the reheating potential. The potential $\potentialthree$ gives the bare potential of $\pithree$, while $\couplingpotential$ generates an alternative local minimum for $\pithree$ (recall that $v$ is negative) when $\pitwo$ approaches to its origin. Thus $\varphi_3$ is moved to this new minimum at $\varphi_3 = 1$ during the \ac{nec} violation. This process is further discussed in \cref{sec:Reheating}.
\item The overall factor introduced for $F_1$ and $F_2$ in eqs.~\eqref{eq:coupling_phi23} modulate the kinetic term of $\pitwo$, which takes $1$ for $\varphi_3 = 0$ and the smaller value $(1 +\alpha_{\rm kick})^{-1}$ for $\varphi_3 = 1$, while $\pithree$ is given an extra \enquote{kick} through $\betakick$ around the time of \ac{nec} violation.
\item The function $G_3$ is taken independent of $\varphi_3$ and makes important contribution only during the \ac{nec}-violating phase.
It returns to a negligible value afterwards, respecting the late-time constraints \cite{Creminelli:2019kjy}.
\end{itemize}

To summarize everything, our total action of \cref{eq:action_pert} consists of the functions $K$, $G_3$ and $P$, whose explicit forms are now
\begin{subequations}
  \label{eq:concrete_forms}
\begin{align}
  K & = \Mpl^2 H_{\rm dip}^2 \left[ \frac{1 + \alpha_{\rm kick} \, {\rm e}^{- \beta_{\rm kick} \varphi_3}}{1 + \alpha_{\rm kick}} \left( F_1(\pitwo) \Xtwo + F_2(\pitwo) \Xtwo^2 \right)
      - {\rm e}^{- \beta_{\rm dip}^2 \left( \left( \varphi_3 - 1 \right)^2 - 1 \right)} v(\pitwo) \right] \; ,
  \label{eq:concrete_K}\\
  G_3 & = \Mpl^2 H_{\rm necv}(\pitwo) \, F_{\rm kb}(\pitwo) \Xtwo \; ,
  \label{eq:concrete_G3}\\
  P & = \Mpl^2 \beta_{\rm kin}^2 \, \Xthree - U(\varphi_3) \; ,
      \label{eq:concrete_P}
\end{align}
\end{subequations}
where $v$ and $F_{\rm kb}$ are given by \cref{eq:tilde_KG3}, $H_{\rm necv}$ by \cref{eq:Hnecv}, $F_1$ and $F_2$ by \cref{eq:F1F2}, and $U$ by \cref{eq:form_P}.
In the following subsection, we discuss the behavior of the effective potential for $\varphi_3$ in more detail.

\subsection{The reheating potential}
\label{sec:Reheating}

We have determined our full Lagrangian as in \cref{eq:action_pert,eq:concrete_forms}. The reconstruction procedure of \cref{subsec:reconst_necv} is robust for the \ac{nec}-violating sector $\pitwo$ as shown in \cref{sec:Stability_and_attractor}, while the reintroduction of the reheating field $\varphi_3$ is done in \cref{subsec:recover_reheating} in a rather \adhoc{} manner.
In this subsection we further describe the trajectory that $\varphi_3$ is expected to take, which shall be verified numerically in \cref{sec:Numerical_approach}. The reheating process operates essentially by the reheating field $\varphi_3$ rolling down to the minimum of its effective potential that is uplifted by the \ac{nec} violation. Then $\varphi_3$ starts oscillating, releasing its energy into the Universe. A coupling of $\varphi_3$ with the Standard Model particles is necessary, but implicit hereafter, and we do not specify its concrete form.

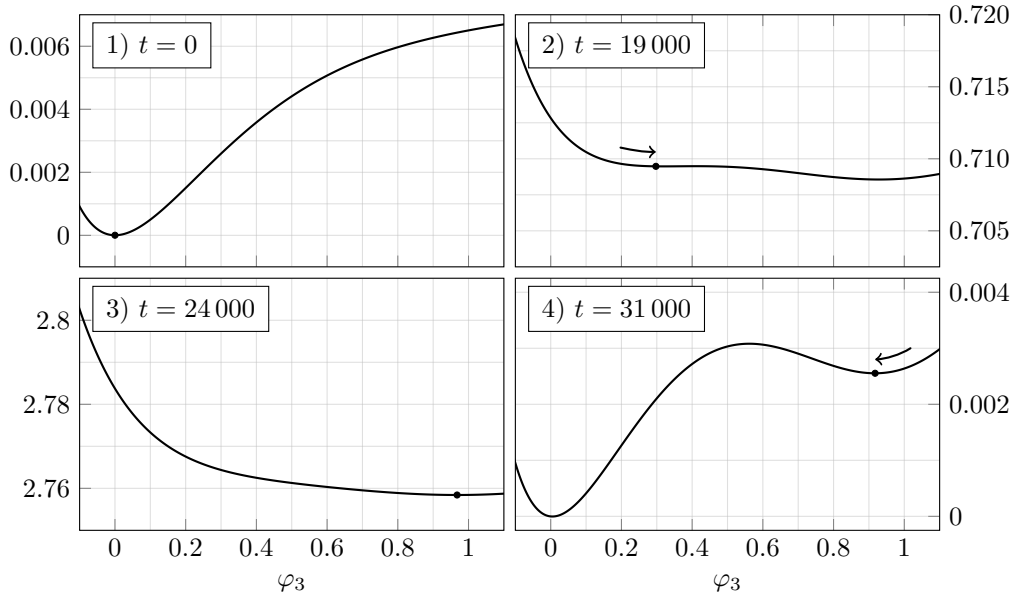
\begin{figure}
    \centering
    \tikzsetnextfilename{reheating_potential_in_multiple_frames}
\begin{tikzpicture}
    %TODO Add arrows
    %TODO Add point giving the field position
    %TODO Add numbers
    %TODO Add legend
    \begin{groupplot}[
        group style={
            group size = 2 by 2,
            vertical sep = 1ex,
            horizontal sep = 1ex,
        },
        grid=both,
        minor tick num = 1,
        minor tick style={draw=none},
        width=0.45\textwidth,
        height=0.2\textheight,
        xmajorgrids={true},
        xmin={-0.1},
        xmax={1.1},
        legend pos = north west,
    ]
        \nextgroupplot[
            tickpos=left,
            xticklabels=none,
            scaled ticks=false,
            tick label style={
                /pgf/number format/fixed,
                /pgf/number format/precision=3
            },
            ymin=-0.001,
            ymax=0.007
        ]

            % t=0
            \addplot[
                color=defaultcolor,
                thick,
                forget plot,
            ] table[col sep=comma,header=false,x index=0,y index=1] {CSVs/rehPotNumWoTimeM2b.csv};

            \filldraw[black] (axis cs:0., 3.01548e-6) circle (0.4mm);

            \addlegendimage{empty legend}
            \addlegendentry{1) $t=\num{0}$};

        \nextgroupplot[
            ytick pos = right,
            xticklabels=none,
            scaled ticks=false,
            tick label style={
                /pgf/number format/fixed,
                /pgf/number format/precision=3,
                /pgf/number format/fixed zerofill,
            },
            ymin=0.7025,
            ymax=0.72
        ]

            % t=19000
            \addplot[
                color=defaultcolor,
                thick,
                % dashed,
                forget plot
            ] table[col sep=comma,header=false,x index=0,y index=2] {CSVs/rehPotNumWoTimeM2b.csv};

            \filldraw[black] (axis cs:0.297442, 0.709477) circle (0.4mm);

            \draw[
                ->,
                thick,
                line cap=round,
            ]
                (axis cs:0.197442, 0.710777) .. controls
                (axis cs:0.237442, 0.710577) and
                (axis cs:0.257442, 0.710477) ..
                (axis cs:0.297442, 0.710477);

            \addlegendimage{empty legend}
            \addlegendentry{2) $t=\num{19000}$};

        \nextgroupplot[
            tickpos=left,
            xlabel=$\pithree$,
            scaled ticks=false, tick label style={/pgf/number format/fixed},
            ymin=2.75,
            ymax=2.81,
        ]

            % t=24000
            \addplot[
                color=defaultcolor,
                thick,
                forget plot,
            ] table[col sep=comma,header=false,x index=0,y index=4] {CSVs/rehPotNumWoTimeM2b.csv};

            \filldraw[black] (axis cs:0.96767, 2.75841) circle (0.4mm);

            \addlegendimage{empty legend}
            \addlegendentry{3) $t=\num{24000}$};

        \nextgroupplot[
            ytick pos = right,
            xlabel=$\pithree$,
            ymin=-0.00025,
            ymax = 0.00425,
            scaled ticks=false,
            tick label style={
                /pgf/number format/fixed,
                /pgf/number format/precision=3
            },
        ]

            % t=31000
            \addplot[
                color=defaultcolor,
                thick,
                forget plot,
            ] table[col sep=comma,header=false,x index=0,y index=6] {CSVs/rehPotNumWoTimeM2b.csv};

            \filldraw[black] (axis cs:0.917477, 0.00255369) circle (0.4mm);

            \draw[
                ->,
                thick,
                line cap=round,
            ]
                (axis cs:1.017477, 0.00300) .. controls
                (axis cs:0.987477, 0.00288) and
                (axis cs:0.947477, 0.00281) ..
                (axis cs:0.917477, 0.00280);

            \addlegendimage{empty legend}
            \addlegendentry{4) $t=\num{31000}$};
    \end{groupplot}
\end{tikzpicture}
    \caption{
      Evolution of $\pithree$ and the shape of its total effective potential, which is altered by $\pitwo$.
      (1) The field $\pithree$ starts here at zero, and (2) it is then raised, until it starts rolling down to (3) a new minimum.
      (4) As the potential regains its previous shape, the field $\pithree$ falls back into (1) its original position.
      We also call for attention to how the vertical scale changes.
    }
    \label{fig:Reheating_potential_schematic}
\end{figure}

The time evolution of the effective potential of $\varphi_3$ is illustrated in \cref{fig:Reheating_potential_schematic}.
The reheating procedure is essentially articulated around the evolution of $\pithree$ along its total potential whose shape is in turn driven by $\pitwo$ in the following way.
%%%%%%%%%%%%%%%%%%%%%%
%%%%%%%%%%%%%%%%%%%%%%
%%%%%%%%%%%%%%%%%%%%%%
%%%%%%%%%%%%%%%%%%%%%%
For most of the cosmic period in which the shift symmetry of $\pitwo$ is well respected, the reheating field $\varphi_3$ stays at the minimum of its bare potential $U$. Once the \ac{nec} violation takes place by $\pitwo$, however, the reheating field $\pithree$ is transported upward and rolls to a new and higher minimum.
Later toward the end of the \ac{nec}-violating phase, the interaction between $\pitwo$ and $\varphi_3$ becomes ineffective, and $\varphi_3$ is subsequently sent back to the initial minimum.
On this return trajectory, the acquired kinetic energy results in oscillations around this minimum, leading to reheating of the Universe.

The shape of the bare potential $U(\varphi_3)$ is of the type of Starobinsky's inflation and is depicted in \cref{fig:Reheating_potential_schematic} (1). The total, \enquote{effective} potential for $\varphi_3$ consists not only of $U(\varphi_3)$ but of the contributions from $K(\pitwo,\Xtwo,\varphi_3)$, as is given in \cref{eq:concrete_K}. A rough description of each term in $K$ is as follows: while each term is approximately constant and takes a negligible value during the (long) \ac{nec}-preserving era, the term proportional to $F_1 \Xtwo + F_2 \Xtwo^2$ is uplifted and erases the minimum around $\varphi_3 = 0$ during the \ac{nec} violation. The term proportional to $v$, on the other hand, arises and accommodates a new minimum at $\varphi_3 = 1$. The changing effective potential is visualized in \cref{fig:Reheating_potential_schematic} (2) and (3). Toward the end of the \ac{nec}-violation period, the modulation by $\pitwo$ effectively turns off, and $\varphi_3$ is drifted back to $\varphi_3 = 0$, as shown in \cref{fig:Reheating_potential_schematic} (4).

The period during the \ac{nec} violation and before the reheating may as well accommodate inflation to produce the seeds of the structure formation. This is also an interesting possibility to realize, but is beyond the scope of our current work, and we leave it to future studies.

%%%%%%%%%%%%%%%%%%%%%%%%%%%%%%%%%%%%%%%%%%%%%%%%%%%%%%%%%%%%%%%%%%%%%%%%%%%%
%%%%%%%%%%%%%%%%%%%%%%%%%%%%%%%%%%%%%%%%%%%%%%%%%%%%%%%%%%%%%%%%%%%%%%%%%%%%

\section{Numerical approach}
\label{sec:Numerical_approach}

Now that we have constructed all the ingredients to achieve the scenario of our interest, we proceed to its numerical verification in this section. Our goals are two-fold: to provide a concrete example in which our model can indeed violate the \ac{nec} in a stable manner, and to demonstrate that a successful reheating follows the \ac{nec}-violating period. We focus on the \ac{nec}-violating-reheating transitions in this work and do not include the cosmological constant relaxation sector $\varphi_1$ in our numerical computation, whose time scale is much longer than the former eras and would thus be computationally rather impractical. As far as the relaxation sector is concerned, we impose the condition of \cref{eq:cond_overshoot_phi2} to ensure that the domination of $\pitwo$ would not lead the motion of $\pione$ to overshoot.

\subsection{Implementation details}

To execute the numerical integration of our model, we solve simultaneously the three equations of motion, \ie.~\cref{eq:EoM_wrt_A,eq:EoM_wrt_pitwo,eq:EoM_wrt_pithree}.
We use the constraint equation, \cref{eq:EoM_wrt_NN}, 
to monitor the numerical convergence, as it must be compatible with the time evolution of the system.
The second derivatives enter the equations only linearly and are hence single-valued at a given time, and the system can be numerically solved using any standard integration algorithm.

The functions $K$, $G_3$ and $P$, appearing in the equations of motion, are substituted by \cref{eq:concrete_forms} together with $v$, $F_{\rm kb}$, $H_{\rm necv}$ and $U$ in \cref{eq:Ktilde,eq:G3tilde,eq:Hnecv,eq:form_P}, respectively, and $\Fone$ and $\Ftwo$ reconstructed as in \cref{eq:F1F2}.
In order to fix a reasonable set of model parameters, we choose the values compiled in \cref{tab:Numerical_parameters} for our numerical integration.
%%%%%%%%%%%%%%%%%%%%%%%%%%%%%%%%%%%%%%
%%%%%%%%%%%%%%%%%%%%%%%%%%%%%%%%%%%%%%
\begin{table}[t]
    \centering
    \begin{tabular}{cccccccccc}
        \toprule
        \multicolumn{2}{c}{Function $H_{\rm necv}$} & \multicolumn{2}{c}{Function $\Fkb$} & \multicolumn{2}{c}{Function $v$} & \multicolumn{2}{c}{Reheating $\pithree$ modulation} & \multicolumn{2}{c}{Reheating $\varphi_3$ sector} \\
        \cmidrule(lr){1-2}\cmidrule(lr){3-4}\cmidrule(lr){5-6} \cmidrule(lr){7-8} \cmidrule(lr){9-10}
        $\quad \Hzero$ & $10^{-3}$ & $\quad \Fkbzero$ & $10^{-3}$ & $\quad v_0$ & $5 \cdot 10^{-2}$ & $\quad \alphakick$ & $10^{-2}$ & $\quad \beta_{\rm kin}$ & $1$
          \\
        $\quad \Hone$  & $1$ & $\quad \Fkbone$ & $1$ & $\quad T_{\rm dip}$ & $2T$ & $\quad \betakick$ & $5$ & $\quad \betaI$ & $3$ \\
        $\quad \HT$ & $5000$ & $\quad \Tkb$ & $3 \HT$ & $\quad H_{\rm dip}$ & $\frac{4 H_1}{10} \sqrt{\frac{\alpha_{\rm kick}}{1 + \alpha_{\rm kick}}}$ &  $\quad \beta_{\rm dip}$ & $2$ & $\quad H_I$ & $\frac{5 H_1}{10} \sqrt{\frac{\alpha_{\rm kick}}{1 + \alpha_{\rm kick}}}$ \\
      $\quad \Htau$  & $500$ &
          & & & \\
        \bottomrule
    \end{tabular}
    \caption{Compilation of all the numerical parameters used in the numerical computations. 
    \label{tab:Numerical_parameters}
    }
\end{table}
%%%%%%%%%%%%%%%%%%%%%%%%%%%%%%%%%%%%%%
%%%%%%%%%%%%%%%%%%%%%%%%%%%%%%%%%%%%%%
The duration of the \ac{nec}-violating period relates to a few parameters, \ie. $T$, $T_{\rm kb}$ and $T_{\rm dip}$, and we thus take them at roughly the same order of magnitude. The ending time of the \ac{nec} violation is controlled by $\tau$, and since we demand the reheating field $\varphi_3$ transit back to its true potential minimum quickly enough to start oscillating, the value of $\tau$ is taken smaller than $T$. The parameter $H_1$ approximately normalizes the value of the Hubble expansion rate at the \ac{nec}-violating era, and we use it as the units for inverse time $t^{-1}$, amounting to fixing $H_1 =1$. In seeking a solution of the approximate form $\partial_t \pitwo \sim 1$, we also measure $\pitwo$ in the units of $H_1^{-1}$. In order for $\pitwo$ to kinetically dominate the energy density during the \ac{nec}-violation, we take the values of $H_{\rm dip}$ and $H_I$ smaller than $H_1$. During the \ac{nec} violation the $\pitwo$ field receives a small \enquote{kick} associated with $\alpha_{\rm kick}$ by the change of $\varphi_3$.
The value of $H_0$ is that of the Hubble expansion rate during the long-lasting \ac{nec}-preserving era, which is to our concern essentially the present Hubble value. It is thus supposed to be extremely small; however, such a huge hierarchy is rather difficult to handle in numerical computations, and we here take a relatively large (but much smaller than $H_1$) value $H_0 = 10^{-3}$ for the purpose of demonstration. Values of order unity are chosen for other parameters.
The reduced Planck mass $\Mpl$ does no affect the dynamics.
Let us emphasize that this set of parameters are chosen to support a proof-of-concept, and other choices can show behaviors similar to the current study.

We take the initial time of the numerical integration  well before the onset of the \ac{nec} violation. We choose to set the initial conditions as
\begin{align}
  \pitwo & = -5 \HT \; , & \pitwodot & = 1 \; , & \pithree & = 0 \; , & \pithreedot & = 0 \; .
\end{align}
That is, the \ac{nec}-violation field $\pitwo$ is assumed to be on its attractor solution, and our choice of $\varphi_3$ potential sets the minimum of $\varphi_3$ at the origin during the period of conserved \ac{nec}. Now all the ingredients are ready to perform the numerical computation.

\subsection{Numerical results}

%%%%%%%%%%%%%%%%%%%%%%%%%%
%%%%%%%%%%%%%%%%%%%%%%%%%%
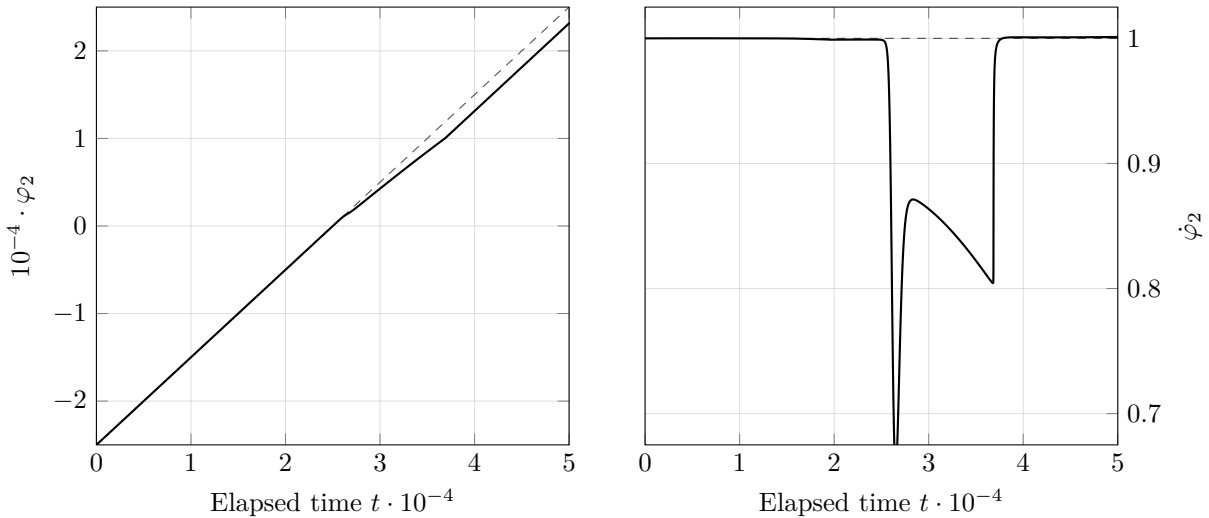
\begin{figure}
  \centering
  \tikzsetnextfilename{Numerical_pitwo_and_pitwodot}
\begin{tikzpicture}[]
    \begin{groupplot}[
        group style={
            group size = 2 by 1,
        },
        width={0.49\textwidth},
        height={0.3\textheight},
        xlabel={Elapsed time $t \cdot 10^{-4}$},
        xmajorgrids={true},
        xmin=0,
        xmax=5,
    ]
        \nextgroupplot[
            ylabel={$10^{-4}\cdot\pitwo$},
            ymajorgrids={true},
            ymin={-2.5},
            ymax={2.5},
        ]
            \addplot[
                color=colorpitwo,
                dashed,
                opacity = 0.7,
                thin,
                domain=-0:5,
                samples=2,
            ] {x-2.5};
            \addplot[
                color=colorpitwo,
                thick,
            ] table[col sep=comma,header=false,x expr=\thisrowno{0}/1e4,y expr=\thisrowno{1}/1e4,] {CSVs/all_together.csv};
        \nextgroupplot[
            ylabel={$\pitwodot$},
            ymajorgrids={true},
            ymin={0.675},
            ymax={1.025},
            ytick pos = right,
        ]
            \addplot[
                color=colorpitwodot,
                thick,
            ] table[col sep=comma,header=false,x expr=\thisrowno{0}/1e4,y index=2] {CSVs/all_together.csv};
            \addplot[
                color=colorpitwodot,
                dashed,
                opacity = 0.7,
                thin,
                domain=-0:5,
                samples=2,
            ] {1.0};
    \end{groupplot}
\end{tikzpicture}
  \caption{
    Time evolution of the \ac{nec}-violating field $\pitwo$ on the left panel, and its time derivative $\pitwodot$ on the right. The dashed line on the left corresponds to $\pitwo = t$, which would be the solution in the absence of $\varphi_3$.}
  \label{fig:Numerical_phi_and_phi_dot}
\end{figure}
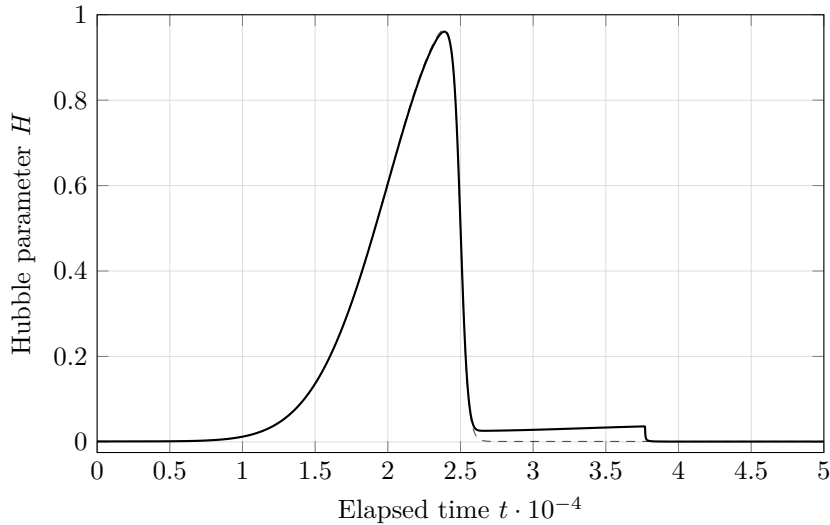
\begin{figure}
  \centering
  \tikzsetnextfilename{Numerical_Hubble}
\begin{tikzpicture}[]
    \begin{axis}[
        width={0.7\textwidth},
        height={0.3\textheight},
        xlabel={Elapsed time $t \cdot 10^{-4}$},
        xmajorgrids={true},
        xmin=0,
        xmax=5,
        ylabel={Hubble parameter $\Hubble$},
        ymajorgrids={true},
        ymin={-0.025},
        ymax={1},
    ]
        \addplot[
            color=colorhubble,
            opacity=0.7,
            dashed,
            thin,
        ] table[col sep=comma,header=false, x expr=\thisrowno{0}/1e4, y index=1] {CSVs/Hguess.csv};
        \addplot[
            color=colorhubble,
            thick,
        ] table[col sep=comma,header=false,x expr=\thisrowno{0}/1e4,y index=5] {CSVs/all_together.csv};
    \end{axis}
\end{tikzpicture}%
  \caption{Time evolution of the Hubble expansion rate $\Hubble$. Notice the overall bell shape as well as the small drop at later times. The input shape $H_{\rm necv}$ (\cref{fig:Hinput_overall_shape}) is here reminded by the dashed line, which would be the behavior of $H$ in the absence of $\varphi_3$.}
  \label{fig:Numerical_Hubble}
\end{figure}
\begin{figure}
  \centering
  \tikzsetnextfilename{Numerical_pithree_and_pithreedot}
\begin{tikzpicture}[]
    \begin{groupplot}[
        width={0.49\textwidth},
        height={0.3\textheight},
        xlabel={Elapsed time $t \cdot 10^{-4}$},
        xmajorgrids={true},
        xmin=0,
        xmax=5,
        group style={
            group size = 2 by 1,
        },
    ]
        \nextgroupplot[
            ylabel={$\pithree$},
            ymajorgrids={true},
            ymin={-0.1},
            ymax={1},
        ]
            \addplot[
                color=colorpithree,
                thick,
            ] table[col sep=comma,header=false,x expr=\thisrowno{0}/1e4,y index=3] {CSVs/all_together.csv};
        \nextgroupplot[
            ylabel={$10^{2}\cdot\pithreedot$},
            ymajorgrids={true},
            ymin={-5},
            ymax={5},
            ytick pos = right,
        ]
            \addplot[
                color=colorpithreedot,
                thick,
            ] table[col sep=comma,header=false,x expr=\thisrowno{0}/1e4,y expr=\thisrowno{4}*1e2] {CSVs/all_together.csv};
    \end{groupplot}
\end{tikzpicture}
  \caption{
    Time evolution of the reheating field $\pithree$ on the left, and its time derivative $\pithreedot$ on the right.
    One can refer to \cref{fig:Reheating_potential_schematic} to observe that $\pithree$ overall follows the target trajectory, and ends up oscillating on its return to the true minimum of the reheating potential.
    The latter oscillation is also clearly visible in the time derivative of $\pithree$, as well as the comparatively small initial move to the temporary equilibrium position.
    }
  \label{fig:Numerical_chi_and_chi_dot}
\end{figure}
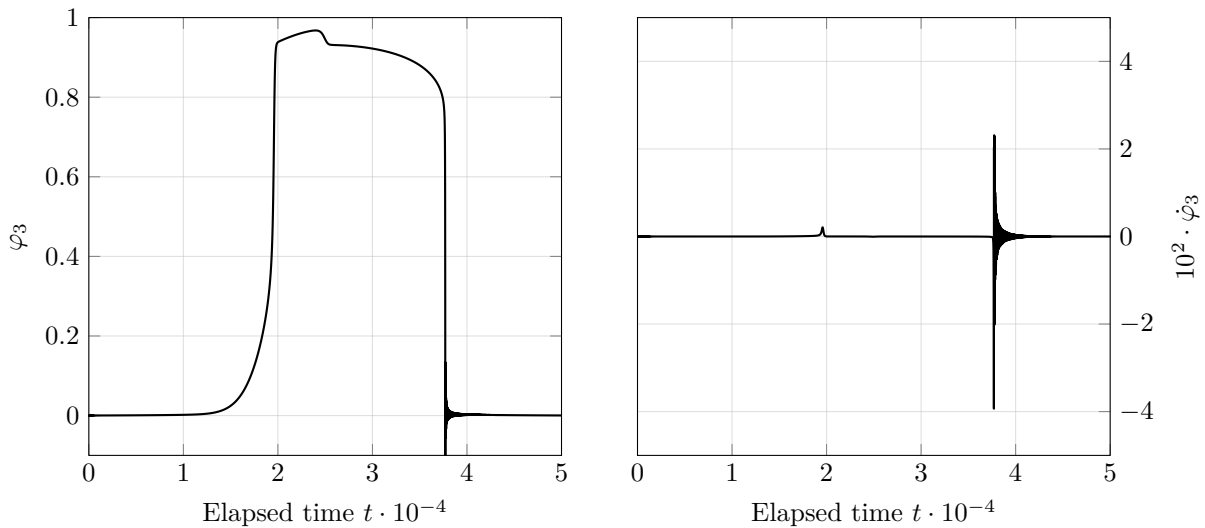

%%%%%%%%%%%%%%%%%%%%%%%%%%
%%%%%%%%%%%%%%%%%%%%%%%%%%
We show the result plots in \cref{fig:Numerical_phi_and_phi_dot,fig:Numerical_Hubble,fig:Numerical_chi_and_chi_dot}.
The first, \cref{fig:Numerical_phi_and_phi_dot}, depicts the trajectories for $\pitwo$ and its time derivative $\dot\varphi_2$. As can be observed, the overall behavior is close to $\pitwo = t$, in accordance to the input \cref{eq:phi2_solution} of the reconstruction process, with a transient deviation during the \ac{nec}-violating phase. This indicates that $\pitwo$ stays closely in the attractor regime discussed in \cref{sec:Stability_and_attractor}, despite the additional ingredient for reheating, which is $\varphi_3$. This justifies our separate treatment of the \ac{nec}-violation and reheating sectors in \cref{sec:A_concrete_implementation} and confirms the validity of our scenario.

%%%%%%%%%%%%%%%%%%%%%%%%%%
%%%%%%%%%%%%%%%%%%%%%%%%%%
%%%%%%%%%%%%%%%%%%%%%%%%%%
%%%%%%%%%%%%%%%%%%%%%%%%%%

The evolution of the Hubble expansion rate, shown in \cref{fig:Numerical_Hubble}, also verifies the desired behavior by violating the \ac{nec} violation for a finite duration. While the value of $H$ drops toward the end of the NECV period around $t \approx 2.5 \cdot 10^4$, it does not reach its final value immediately, but instead it stays at a larger value for some time until finally dropping to the NEC-preserving value $H_0$. This is due to the contribution from the reheating field $\varphi_3$, which is trapped at the local minimum of its effective potential as illustrated in the bottom panels of \cref{fig:Reheating_potential_schematic}.
Except for this small modulation, the overall behavior mimics the shape of the predetermined $H_{\rm necv}$ in \cref{eq:Hnecv}, which is co-drawn as a dashed curve in \cref{fig:Numerical_Hubble}.

%%%%%%%%%%%%%%%%%%%%%%%
%%%%%%%%%%%%%%%%%%%%%%%
%%%%%%%%%%%%%%%%%%%%%%%
%%%%%%%%%%%%%%%%%%%%%%%
Thanks to the \ac{nec} violation, $\pitwo$ effectively stores energy, and the field $\pithree$ that is coupled to it in turn acquires part of the energy to reheat the Universe. For successful reheating and efficient energy release, it is essential for $\varphi_3$ to actually rolls down at the end of \ac{nec} violation and start oscillating once it falls back to its true potential minimum. This is the dynamics indeed observed in the numerical evolution as in \cref{fig:Numerical_chi_and_chi_dot}. As seen, the initially stabilized $\varphi_3$ at $\varphi_3 = 0$ is displaced at the beginning of the \ac{nec}-violating phase through its interaction with $\pitwo$. When $H$ drops around $t \approx 2.5 \cdot 10^4$, $\varphi_3$ is left at the same local minimum, though its value slightly changes due to the drop. It stays there for some time, during which its effective potential regains the true minimum at the origin as illustrated in the bottom right panel of \cref{fig:Reheating_potential_schematic}. Eventually $\varphi_3$ drops toward the minimum and starts the oscillation, which is the moment reheating occurs. This concludes the concrete demonstration of the reheating mechanism of the Universe that would otherwise be empty after the relaxation phase of the cosmological constant.

A few consistency checks are in order. First, we ensure that the presence of the \ac{nec} violation should not mess up the relaxation mechanism, namely the condition for no overshooting, \cref{eq:cond_overshoot_phi2}, needs to be imposed. Under the current parametrization of the numerics, $\partial_t \pitwo \equiv M^2 \approx 1$, and $t$ is measured in the units of $H_1^{-1}$, and so is $\pitwo$, where $H_1$ is approximately equal to the value of the Hubble expansion rate during the \ac{nec} violation. Then the condition of \cref{eq:cond_overshoot_phi2} translates to
\begin{align}
  \label{eq:cond_overshoot_numerics}
  \frac{\Delta \pitwo}{H_1^{-1}} \lesssim \left( \frac{\Mpl}{\sqrt{3} \, H_1} \right)^{4m-2} \, \frac{H_0^2}{\Mpl^2} \; .
\end{align}
In our example numerical computation, $\Delta \pitwo$ for the \ac{nec} violation is about a few times $10^{4}$ in the units of $H_1^{-1}$, and if we take $H_1 \sim 10^{12} \, {\rm GeV} \sim 10^{-6} \Mpl$, assuming a high \ac{nec} violating energy scale, we have $H_0 = 10^{-3} H_1 \sim 10^{-9} \Mpl$. Then the above condition of \cref{eq:cond_overshoot_numerics} is satisfied for $m \gtrsim 1.45$, which is weaker than the theoretical constraint $m>3/2$ already imposed (see after \cref{eq:V1_approaches_0_when_N_goes_to_infinity}). Note that the realistic value of $H_0$ is much smaller than that taken in our numerical example, and nonetheless the condition can be met for a larger value of $m$ without difficulty.

%%%%%%%%%%%%%%%%%%%%%%%%%%%%%%
%%%%%%%%%%%%%%%%%%%%%%%%%%%%%%
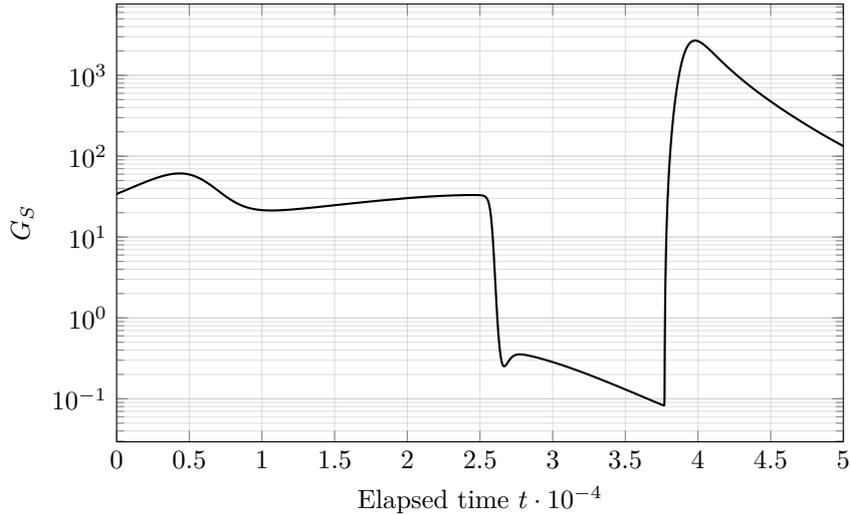
\begin{figure}
  \centering
  \tikzsetnextfilename{GS}
\begin{tikzpicture}[]
    \begin{axis}[
        height=0.3\textheight,
        width=0.7\textwidth,
        xmajorgrids={true},
        xmin=0,
        xmax=5,
        ymode = log,
        xlabel={Elapsed time $t \cdot 10^{-4}$},
        ylabel={$G_S$},
        ymajorgrids={true},
        ]
        \addplot[
            color=Black,
            thick,
        ] table[col sep=comma,header=false,x expr=\thisrowno{0}/1e4,y index=1] {CSVs/GS.csv};
\end{axis}
\end{tikzpicture}
  \caption{Time evolution of the no-ghost condition $\mathcal{G}_S$, defined in \cref{eq:matrix_T}. Its value is observed to remain positive, satisfying the first no-ghost condition in \cref{eq:noghost}. Note that the other no-ghost condition is trivially satisfied, see the main text.}
  \label{fig:GS}
\end{figure}

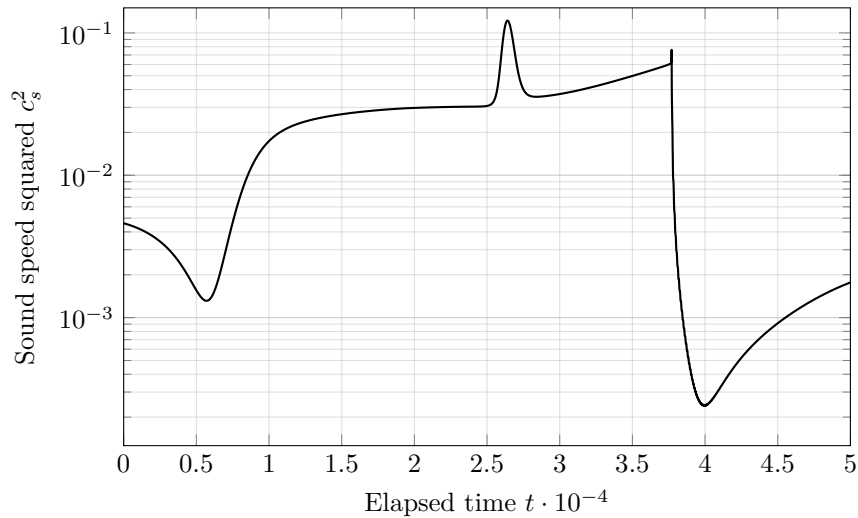
\begin{figure}
    \centering
    \tikzsetnextfilename{Numerical_cstwo}
\begin{tikzpicture}[]
    \begin{axis}[
        width={0.7\textwidth},
        height={0.3\textheight},
        xlabel={Elapsed time $t \cdot 10^{-4}$},
        xmajorgrids={true},
        xmin=0,
        xmax=5,
        ylabel={Sound speed squared $\cstwo$},
        ymajorgrids={true},
        ymax={0.15},
        ymode=log,
    ]
    \addplot[
        color=colorcstwo,
        thick,
    ] table[col sep=comma,header=false,x expr=\thisrowno{0}/1e4,y index=6] {CSVs/all_together.csv};
\end{axis}
\end{tikzpicture}
    \caption{
        The square of the speed of sound, $\cstwo$, remains strictly positive and below unity during the whole cosmic history, showing the absence of gradient instability and superluminality. The other sound speed is trivially $1$ in our present case.}
    \label{fig:Numerical_cs2}
\end{figure}
%%%%%%%%%%%%%%%%%%%%%%%%%%%%%%
%%%%%%%%%%%%%%%%%%%%%%%%%%%%%%
Moreover, we can easily verify that the stability conditions against small perturbations are also fulfilled in our result. Among the no-ghost conditions summarized in \cref{eq:noghost}, the second one, $\mathcal{G}_\varphi > 0$, is trivially satisfied with our choice $P \propto \Xthree$ (\ie. $\mathcal{G}_\varphi = \MPlSq \betakin^2$), and the first one, $\mathcal{G}_S >0$, is numerically confirmed in \cref{fig:GS}.
On the other hand, the squared sound speeds $c_s^2$ obtained in \cref{eq:cs2_simple} do not invoke any instabilities. One of the values is trivially unity, $c_\chi^2 = 1$, in our case, and the other, $c_s^2 = \mathcal{F}_S/\mathcal{G}_S + \mathcal{A}$, is shown to be bounded by $0 < c_s^2 < 1$ in \cref{fig:Numerical_cs2} for the entire duration of the computation, and thus neither gradient instability nor superluminality is present.

The above results therefore show the realization of our target scenario: while $\pione$ operates the relaxation mechanism of cosmological constant, reviewed in \cref{subsec:relax}, the field $\pitwo$ stably violates the \ac{nec}, and the reheating field $\pithree$ moves to a new minimum, before oscillating on its way back to its true minimum and eventually reheating the Universe.

%%%%%%%%%%%%%%%%%%%%%%%%%%%%%%%%%%%%%%%%%%%%%%%%%%%%%%%%%%%%%%%%%%%%%%%%%%%%
%%%%%%%%%%%%%%%%%%%%%%%%%%%%%%%%%%%%%%%%%%%%%%%%%%%%%%%%%%%%%%%%%%%%%%%%%%%%

\section{Conclusion and outlook}
\label{sec:Conclusion_and_outlook}

The current work exhibits a concrete model that both answers to the cosmological constant problem by dynamically relaxing the \ac{cc}, and subsequently reheats the Universe (\cref{sec:Overall_picture_of_the_cosmic_history}). It builds upon two previous works, \cite{Mukohyama:2003nw,Mukohyama:2003ac} and \cite{Alberte:2016izw}, and extends the studies on stability and potential overshooting issues with numerical confirmation.
This work provides a conceptual proof of a system that resolves the \ac{cc}~problem without fine-tuning.

Our proposed model is the stable assembly of three components.
Firstly, the model introduces a scalar field $\pione$ \cite{Mukohyama:2003nw,Mukohyama:2003ac} equipped with an atypical kinetic term modulated by an inverse power of the spacetime curvature invariant, which effectively lets the field $\pione$ roll down and eventually has its potential converge to a tiny, but positive value.
This final value becomes an effective \ac{cc}.
This dynamical relaxation process \enquote{feels} the value of any existing contributions to the \ac{cc}, including quantum vacuum energy, and fixes the classical vacuum expectation value of $\varphi_1$ such that it cancels out the \ac{cc}. The mechanism is not vulnerable to radiative corrections to curvature either, thus enjoying the advantage of avoiding known fine-tuning issues associated with \ac{cc}.
However, it also effectively empties the Universe, by diluting its contents such as radiation and matter by the cosmic expansion. 
This is by construction an inevitable consequence from the $\varphi_1$ sector alone, and an additional ingredient is in need for successful cosmology.

Secondly, to resolve this newly introduced issue, 
two other scalar fields $\pitwo$ and $\pithree$ work together to violate the null-energy condition and repopulate our Universe.
The former field $\pitwo$ goes through the dynamics that effectively raises the value of the Hubble expansion rate for transient periods, thus breaking the null-energy condition.
In order to ensure that the \ac{nec} violation does not destroy the \ac{cc}~relaxation mechanism of $\varphi_1$, we impose the condition of \cref{eq:cond_overshoot_phi2} to avoid the case in which $\pitwo$'s motion irreversibly lets $\varphi_1$ overshoot the zero of the effective \ac{cc}. We also require the time scale of the former be much shorter than that of the latter, giving sufficient time for the \ac{cc} relaxation to operate, and that $\pitwo$ respects (approximate) shift symmetry in the regions where the \ac{nec} is preserved. Moreover, for the sake of naturalness with respect to the timing of \ac{nec} violation, the field space of $\pitwo$ is assumed to be periodic with a period much longer than the duration of a single \ac{nec} violation phase. This way our Universe goes through this phase multiple times and our current Universe merely occurs after many cycles of them.
The field $\pithree$ acts as a reheating field that extracts the energy from the \ac{nec} violation sector, starts oscillating after it, and eventually reheats the Universe.

Thirdly, the last component of our model is the gravity sector. It has non-minimal couplings to $\varphi_1$ and $\pitwo$, as already described. As the metric-only part, together with the standard Einstein-Hilbert term, our action includes the quadratic term of the Ricci scalar, which stabilizes the \ac{cc}~relaxation sector \cite{Mukohyama:2003nw,Mukohyama:2003ac}. These three ingredients together achieve our complete cosmological scenario of \ac{cc}~relaxation, \ac{nec} violation, and reheating, in a stable manner.

In order to concretely realize the desired cosmological history, the model functions need to be fixed, and  we proceed to their reconstruction in \cref{sec:A_concrete_implementation}. While we determine the $\pitwo$ sector in an unambiguous manner, the way $\varphi_3$ is included is rather \adhoc. Nevertheless we conduct numerical integration for the entire system of $\pitwo$ and $\varphi_3$ (but without $\varphi_1$, for numerical ease) in \cref{sec:Numerical_approach}, thus justifying our methodology and confirming the realization of \ac{nec} violation followed by a reheating phase. The stability conditions against small perturbations are also shown to be respected.

In our numerical example, the energy used for reheating is roughly one order of magnitude smaller than the energy acquired by the \ac{nec} violation in the units of $H$, as seen in \cref{fig:Numerical_Hubble,fig:Numerical_chi_and_chi_dot}.
This is basically the limitation due of the concrete implementation we have chosen, and we leave to future studies the possibility of more efficient energy transfer and higher reheating scale.
Moreover, our current choice of parameters does not accommodate an inflationary phase and thus has no implications (for now) for the structure formation and cosmic microwave background anisotropies. The connection to these observables is beyond the scope of our present work, as satisfying all the stability conditions simultaneously is already a non-trivial task in the minimal scenario. We would like to return to this issue in the future work.

Our model is based on the Horndeski class of theories to violate the \ac{nec}. As an another possibility, the so-called minimally modified gravity theories \cite{Lin:2017oow,Mukohyama:2019unx,DeFelice:2020eju} could be an interesting candidate. While this class of theory contains only two degrees of freedom and consequently avoids instabilities with no difficulty, it nonetheless allows a large freedom for background evolution, not limited by energy conditions. Our conceptual scenario may thus provide a fertile ground for model buildings.

The concrete model considered in this work is an effective theory. Linking it to some more fundamental UV theory could also lead to an interesting path to explore. That would accomplish a complete, self-containing solution to the long-standing \ac{cc}~problem. We wish our present study gives one step forward in healing this conceptual pathology.

%%%%%%%%%%%%%%%%%%%%%%%%%%%%%%%%%%%%%%%%%%%%%%%%%%%%%%%%%%%%%%%%%%%%%%%%%%%%

\bibliographystyle{unsrt}
\bibliography{bibliography}
\addcontentsline{toc}{section}{References}

\appendix

\end{document}